\documentclass[11pt,notitlepage,longbibliography]{revtex4-1} 

\usepackage{amsmath,amssymb,mathtools,mathrsfs}
\usepackage[colorlinks=true,citecolor=blue,linkcolor=blue,urlcolor=blue]{hyperref}%
\usepackage[usenames,dvipsnames]{xcolor}

\usepackage{ulem}
\newcommand{\AvdW}{\mathcal{V}_\gamma}

\newcommand{\dd}{\mathrm{d}}
\newcommand{\eH}{\tilde{H}}
\newcommand{\eX}{\tilde{X}}
\newcommand{\ePi}{\tilde{\Pi}}
\newcommand{\eT}{\tilde{T}}
\newcommand{\ind}{{\bf 1}_{\mathrm{drop}}}

\newcommand{\hinf}{h_\infty}
\newcommand{\Hinf}{H_\infty}
\newcommand{\heqm}{h_\mathrm{eqm}}
\newcommand{\Heqm}{H_\mathrm{eqm}}
\newcommand{\xmin}{x_{\mathrm{min}}}
\newcommand{\Xmin}{X_{\mathrm{min}}}
\newcommand{\Hmin}{H_{\mathrm{min}}}
\newcommand{\Xmax}{X_{\mathrm{max}}}
\newcommand{\Hmax}{H_{\mathrm{max}}}
\newcommand{\xinf}{x_\infty}
\newcommand{\Xinf}{X_\infty}
\newcommand{\Pinf}{P_\infty}
\newcommand{\lc}{\ell_c}
\newcommand{\lmen}{\ell_{\mathrm{men}}}
\newcommand{\Lmen}{L_{\mathrm{men}}}

\newcommand{\Tgrav}{T_\mathrm{grav}}
\newcommand{\Grav}{\mathcal{G}}

\newcommand{\Bo}{\mathrm{Bo}}
\newcommand{\Tplateau}{T_{\mathrm{plat}}}
\newcommand{\Hplateau}{H_{\mathrm{plat}}}
\newcommand{\As}{\alpha}
\newcommand{\dt}{\Delta\theta}
\newcommand{\pdrop}{p_{\mathrm{drop}}}

\newcommand{\Vskirt}{V_\mathrm{skirt}}
\newcommand{\vskirt}{v_\mathrm{skirt}}
\newcommand{\Vdrop}{V_\mathrm{drop}}
\DeclareMathOperator{\sech}{sech}

\begin{document}

\title{Droplets on Lubricated Surfaces: The slow dynamics of skirt formation}
\author{Zhaohe Dai and Dominic Vella}
 \email{dominic.vella@maths.ox.ac.uk}
\address{Mathematical Institute, University of Oxford, Woodstock Rd, Oxford, OX2 6GG, UK}
\date{\today}

\begin{abstract}
A key question in the interaction of droplets with lubricated and liquid-infused surfaces is what determines the apparent contact angle of droplets. Previous work has determined this using measured values of the geometry of the `skirt' --- the meniscus-like deformation that forms around the base of the deposited droplet. Here, we consider theoretically the equilibrium of a droplet on a smooth, impermeable lubricant-coated surface, and argue that the small effect of gravity within the skirt and the size of the substrate are important for determining the final equilibrium. However, we also show that the evolution of the skirt towards this ultimate equilibrium is extremely slow (on the order of days for typical experimental parameter values). We therefore suggest that previous experiments on smooth lubricated  surfaces may have observed only slowly-evolving transients, rather than `true' equilibria, potentially explaining why a wide range of skirt sizes have been reported.
\end{abstract}

\maketitle

\section{Introduction}

Liquid-infused surfaces (LISs), also known as slippery liquid infused porous surfaces (SLIPSs), are formed by coating surface with a thin layer of oil lubricant \cite{quere2005,Wong2011}. This liquid coating provides a barrier that prevents  droplets of other liquids from reaching the solid surface and thereby allows deposited droplets to move with ultra-low friction  \cite{Li2020,Smith2013,Daniel2017,Keiser2017,Keiser2020}. This very low friction has found a variety of applications including the creation of surfaces that are anti-biofouling \cite{Epstein2012}, anti-icing \cite{Kreder2016} and facilitate  water harvesting \cite{Guo2021}, as well as allowing new routes for  droplet manipulation  \cite{Orme2019,Jiang2019,Hack2018}.

In many of these applications,  the deformation of the lubricant surface caused by a droplet is particularly important \cite{Hardt2021}; for example, meniscus-like deformations around the base of a droplet, often called a `skirt' \cite{Schellenberger2015,Villegas2019,Peppou2020}, give rise to interactions between droplets \cite{Hack2018,Jiang2019,Orme2019} that is a droplet analogue of the ``Cheerios effect'' \cite{Vella05}. However, there appears to be no clear consensus on what the size of this skirt region is. For example, Semprebon \emph{et al.}~presented theoretical arguments for the apparent contact angle assuming given (constant) curvatures of the interfaces in the skirt region \cite{Semprebon2017,Semprebon2021}. Their results were consistent with some experiments, in which the radius of curvature of the skirt was constant and, further, small compared with the droplet size \cite{Kreder2018,Daniel2017,Keiser2020,Mchale2019}. However, in other experiments, \citet{Schellenberger2015} observed menisci that did not have constant curvature  and were of a size comparable to the capillary length (and hence significantly larger than the droplet). Even without this discrepancy, current models appear unable to predict either the height of the triple  line (along which droplet, lubricant and vapor meet) in the equilibrium state or the effective contact angle; for example, the theory of \citet{Semprebon2021} takes the ratio of pressures within the droplet and layer as a control parameter, but does not determine from first principles what this ratio should be.


In this paper, we consider the formation and ultimate equilibrium of the skirt region formed when a droplet is deposited on the surface of a LIS. Our aim is to present results in terms of fundamental geometrical and physical control parameters of the system, rather than emergent properties. While many implementations of LIS use a microscopic texture to retain the lubricant layer within the texture, we consider the (simpler) case of an impermeable, smooth substrate coated by a layer of oil, often referred to as a lubricated surface \cite{Schellenberger2015,Daniel2017,Kreder2018,Scarratt2020}.


 For the case of  lubricated surface, the surface remains wetted by a thin oil layer throughout and two key questions arise: (i) what is the equilibrium state? (ii) how long after deposition is this state observed? These two questions are the focus of this paper. We begin in \S\ref{sec:Setup} by describing a model problem for equilibrium that includes the essence of the forces on a lubricating layer that are induced by the presence of a droplet: the `pulling' caused by the droplet's interfacial tension and the `pushing' caused by its internal capillary pressure. Despite the apparent simplicity of this model problem, in \S\ref{sec:AsyStatics} we derive asymptotic results showing that many disparate length scales enter: the thickness of the oil layer, the size of the substrate and gravity all play some role in selecting the final equilibrium. In particular, we show that the availability of enough lubricant to reach the desired equilibrium is not guaranteed and consider the asymptotic limits that appear when either the droplet is starved of lubricant \cite{Tress2017} or has its need for lubricant sated (in senses to be defined). In \S\ref{sec:Dynamics}, we study the dynamic approach to this equilibrium state. We find that the system passes through many different phases  to reach the final equilibrium and, crucially, find that the true equilibrium is only approached on extraordinarily long time scales. We suggest that this long time scale, as well as the complex role of many different length scales in determining the equilibrium,  may be the reason that different experiments have reported different behaviours. We finish by summarizing our findings and discussing possible refinements and extensions of the model in \S\ref{sec:Conclusions}.

\section{Equilibrium\label{sec:Setup}}

\subsection{Model problem: pushing and pulling a thin film}

The effect of an axisymmetric droplet with contact angle  $\theta$  and triple line radius $R_c$  deposited on a thin oil layer  is two-fold: firstly the capillary pressure within the droplet, $\pdrop=2\gamma_{dv}\sin\theta/R_c$, squeezes oil from beneath the droplet; secondly, the capillary force from the droplet--vapor interface, $2\pi\gamma R_c \sin\theta$, pulls the oil interface upwards, sucking liquid into the wetting skirt as it goes.  (Note that in this discussion the weight of the drop is neglected entirely since the drops of interest are usually `small' in a sense that we quantify in due course.) The squeezing action of the droplet capillary pressure has been appreciated previously by \citet{Daniel2017}, who noted that this pressure is ultimately balanced by the repulsive van der Waals pressure, $p_\mathrm{vdW}\approx A/h^3$ with $A$ the Hamaker constant; this leads to an equilibrium film thickness beneath the droplet  $\heqm\approx (A R_{\mathrm{drop}}/\gamma)^{1/3}$. 

The equilibrium of the triple line has been considered by a variety of authors; perhaps most notably, \citet{Semprebon2017,Semprebon2021} considered the equilibrium of the Neumann triangle at the triple line under the assumption that external contact lines form on the planar substrate. At the same time, the early stages of this formation have been studied in a related problem by \citet{Hack2018}. However, how these early stages connect to the ultimate equilibrium has not, to our knowledge, been studied.

\begin{figure}[htp]
    \centering
    \includegraphics[width=10.5cm]{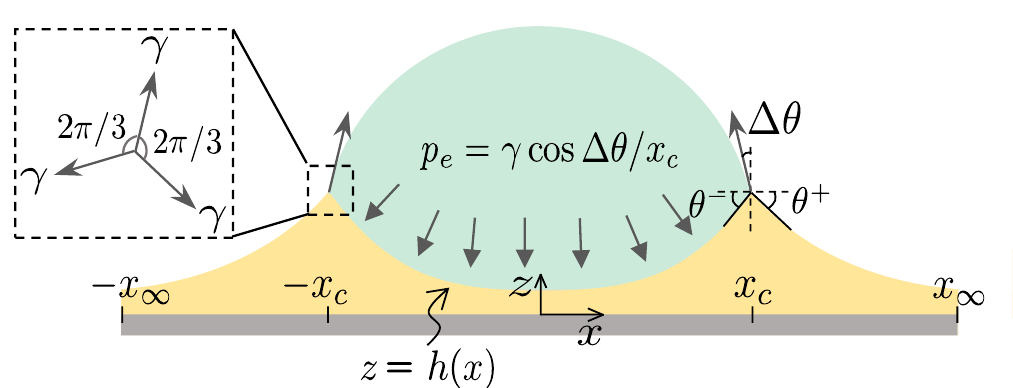}
    \caption{Schematic illustration and notation for the analysis of the `push-and-pull' problem considered in this paper. We mimic the effect of a two-dimensional liquid droplet sitting on a thin oil film by introducing a constant pressure pushing down on the film (mimicking the droplet Laplace pressure) within a constant interval $|x|<x_c$, together with line forces pulling on the interface at $x=\pm x_c$ (mimicking the capillary force acting at the triple lines). This combination of loads leads ultimately to the formation of an equilibrium meniscus, given by $z=h(x)$. The inset figure shows a zoom of the triple line region, which, for the case of equal surface tensions considered here, gives rise to \eqref{eq:3PointStar}.}    \vspace{0.5cm}
    \label{fig:Schematic}
\end{figure}

In this paper, we seek to understand the interaction of the squeezing out of liquid beneath the droplet and the pulling up of the triple line by surface tension. To make this problem more tractable, we consider the model two-dimensional problem shown schematically in fig.~\ref{fig:Schematic}: the droplet is represented by a positive pressure $p$ acting over $|x|\leq x_c$ while the effect of the tension of the droplet--vapour interface is modelled by two line forces, each of magnitude $\gamma$, pulling at an angle $\Delta\theta$ to the vertical at $x=\pm x_c$. (Note that this corresponds to assuming an apparent contact angle $\theta_A=\pi/2-\Delta\theta$ and, further, that $\Delta\theta$ may evolve dynamically as the skirt forms. Moreover, the rotation of the pulling force might also change $x_c$ through the constraint on the droplet's volume; we neglect such complexity in this model problem.) To make the problem more tractable still, we assume that all interfacial tensions are equal, i.e.~$\gamma_{ov}=\gamma_{od}=\gamma_{dv}=\gamma$; this ensures that a triple-line forms since the spreading parameter $S=\gamma_{dv}-\gamma_{ov}-\gamma_{od}=-\gamma<0$ and the droplet cannot be `cloaked' by the liquid \cite{Schellenberger2015}. With this assumption, the Neumann balance (see inset of fig.~\ref{fig:Schematic}) on the triple line immediately gives that
\begin{equation}
    \theta^\pm\pm\dt=\pi/6.
    \label{eq:3PointStar}
\end{equation} (We shall see that in this case, the rotation of the pulling force is moderate even when the skirt is ``large".) Global vertical force balance on the liquid then requires that $p=\gamma(\cos\dt)/x_c$; alternatively, small droplets are circular, with radius of curvature $x_c/\cos\dt$ given by elementary geometry. 

Our primary interest lies in understanding how the combination of the two effects of the droplet described above affects the formation of the skirt that forms around a droplet placed on a lubricating layer. What limits the final size of the skirt? Over what time scale does the skirt develop? How does the apparent contact angle (or the tilting of the Neumann triangle $\Delta\theta$)  evolve away from $\theta_A=\pi/2$ with time? Before addressing questions about the evolution of the surface of the lubricating oil layer, we first turn to describe the equilibrium that this `push-and-pull' system ultimately reaches.

\subsection{Mathematical model}

The two effects of the droplet are introduced into the pressure field that is imposed on the lubricating film using the indicator function, $\ind(x)$, for the droplet region (i.e.~$\ind(x)=1$ for $|x|\leq x_c$ and $\ind(x)=0$ otherwise) to describe the constant pushing pressure of the droplet's interior and a Dirac $\delta$-function to describe the capillary force from the droplet--vapor interface. In particular, we have
\begin{equation}
    p_e(x)=\frac{\gamma\cos\Delta\theta}{x_c}\ind(x)-\gamma\cos\Delta\theta[\delta(x+x_c)+\delta(x-x_c)]
    \label{eq:PressDis}
\end{equation}
for $|x|\le\xinf$, with $2\xinf$ the lateral extent of the thin liquid film.

The pressure within the liquid film is then determined by combining this pressure with the pressure jump due to surface tension, the hydrostatic pressure within the liquid and a contribution from van der Waals forces. We find that
\begin{equation}
    p(x,z)=p_e(x)-\gamma \frac{h_{xx}}{\left(1+h_x^2\right)^{3/2}} - \frac{A}{h^3} + \rho g (h-z),
    \label{eq:Pressure}
\end{equation}
where $A>0$ is the Hamaker constant (so that the minus sign gives a repulsive vdW pressure between the interface and the substrate,  preventing the oil from draining entirely away) and $z$ is the vertical position within the film. Note that \eqref{eq:Pressure} includes the effects of both gravity and van der Waals forces acting on the lubricant film. While these forces usually act at very different scales, we shall see that in this problem they both play an important role. (In particular, van der Waals forces will be important beneath the droplet, while gravity is important in the region beyond the droplet.) We have seen that the equal surface tensions assumed here, combined with the Neumann relations leads to \eqref{eq:3PointStar}; as such, not all of the angles $\theta^\pm$ and $\dt$ can be small, and we must retain the geometrically nonlinear curvature in \eqref{eq:Pressure}, even though the thickness of the oil film is small compared to its horizontal scale, $x_c$.

In equilibrium, there cannot be a horizontal pressure gradient and so, considering the pressure along $z=0$ in particular, we find
\begin{equation}
    p_e(x)-\frac{\gamma h_{xx}}{\left(1+h_x^2\right)^{3/2}} - \frac{A}{h^3} + \rho gh - p_\infty =0,
    \label{eq:static}
\end{equation}
where $p_\infty$ is the pressure at the edge of the plate that is not known \emph{a priori} and must be determined as part of the solution. However, we expect the interface far from the droplet to return to a constant height, $\hinf$ so that the pressure there is
\begin{equation}
    p_\infty=-A/\hinf^3 + \rho g \hinf,
\end{equation} measured relative to the  (atmospheric) pressure datum.  Note that this far-field film height is, in general, different from the initial liquid film height, $h_0$, before the droplet is deposited. Nevertheless, we expect $\hinf\to h_0$ when the system becomes large, i.e.~as $\xinf\to\infty$.

 Equation \eqref{eq:static} is an equation for the meniscus shape $h(x)$ in equilibrium and is to be solved with symmetry boundary conditions at $x=0$ and $\xinf$ as well as the global conservation of volume (which encodes the pertinent information about the initial condition); these conditions read
\begin{equation}
    h_x(0)=h_x(\xinf)=0\quad\text{and}\quad\int_0^{\xinf} h\,\dd x = h_0\xinf.
    \label{eq:staticsBCs}
\end{equation}

In practice, the $\delta$-function singularity in $p_e(x)$ is best handled by integrating \eqref{eq:static} across $x=x_c$, which gives two additional boundary conditions:
\begin{equation}
    h(x_c^-)=h(x_c^+),\quad\text{and}\quad \left. \frac{h_x}{\left(1+h_x^2\right)^{1/2}}\right\vert_{x_c^+}-\left. \frac{h_x}{\left(1+h_x^2\right)^{1/2}}\right\vert_{x_c^-}=-\cos\dt.
    \label{eq:staticsBCs2}
\end{equation}
Physically, the second condition in \eqref{eq:staticsBCs2} ensures that the line forces at the triple line are balanced vertically. It is natural also to ask how the horizontal force balance at the triple line can be satisfied within this model. In short, any asymmetry between the meniscus slopes either side of the triple line  necessarily induces a rotation of the direction along which the droplet's line force acts. We denote this rotation by $\dt$, with the sign convention that $\dt>0$ in the counter-clockwise direction (see fig.~\ref{fig:Schematic}), so that
\begin{equation}
\sin\dt=\cos\theta^+-\cos\theta^-=    \left. \left(1+h_x^2\right)^{-1/2}\right\vert_{x_c^+}-\left. \left(1+h_x^2\right)^{-1/2}\right\vert_{x_c^-}
\label{eqn:CLAsymmetry}
\end{equation} from horizontal force balance, where $\theta^+$ and $\theta^-$ are meniscus angles evaluated on the right and left side of the triple line (see fig.~\ref{fig:Schematic}).

\subsection{Some insights from scaling}

We shall shortly present a detailed study of the solution of the system \eqref{eq:static}--\eqref{eq:staticsBCs}. However, to do this relies on an appropriate non-dimensionalization of the problem, which in turn depends on some understanding of how the system is likely to behave. Given that there are disparate length scales in this problem (in addition to the size of the `droplet', $x_c$, and the initial thickness of the liquid layer) it is helpful to think about this in terms of what the system would `like' to do.

Since the liquid beneath the droplet, $0<x<x_c$, is subject to a positive capillary pressure, magnitude $\gamma/x_c$, it is clear that this interface will seek to be an arc of a circle with radius of curvature $\propto x_c$ in the absence of gravity, i.e.~if $x_c\ll\lc=(\gamma/\rho g)^{1/2}$. In this limiting case, therefore, the profile of the meniscus beneath the droplet $h(x)\propto x^2/x_c$; in particular, the height of the triple line $h_c=h(x_c) \propto x_c$. While the behaviour of the inner meniscus determines the height of the triple line in this limit, the outer meniscus (beyond the triple line) is not subject to a downward pressure, and so can only return to being flat because of the combined effect of the van der Waals attraction and hydrostatic pressure (this explains why both effects must be accounted for). Assuming for the moment that the film thickness is such that hydrostatic pressure dominates van der Waals force throughout, we therefore expect that the outer meniscus should decay over the capillary length $\lc=(\gamma/\rho g)^{1/2}$. (Previous experimental data on inverse opal surfaces confirm this importance of gravity \cite{Schellenberger2015}, as also discussed in Appendix \ref{sec:Experiments}.)  At a scaling level, then, the total volume of liquid lifted into the skirt in this limit is
\begin{equation}
    \vskirt\propto x_c^2+x_c\lc\approx x_c\lc
    \label{eqn:VolScaling}
\end{equation} (since $x_c\ll\lc$ by assumption). 

The volume of liquid trapped within the skirt \eqref{eqn:VolScaling} involves the droplet size and the capillary length, $\lc$, but takes no account of the total volume of lubricating liquid available to the system, which is simply $x_\infty h_0$. Clearly if the reservoir of liquid stored within the lubricating film is sufficiently large the system will be able, at least in principle, to reach its desired equilibrium: the skirt's appetite for lubricant is `sated'. However, if $x_\infty h_0\lesssim x_c\lc$ then the amount of lubricant is limite, the skirt is `starved' of lubricant and the  system must  do something else. Determining this alternative is the ultimate aim of \S\ref{sec:AsyStatics}, but  the insight already gained helps to choose appropriate scales for the non-dimensionalization of the problem.

\subsection{Non-dimensionalization}

The preceding discussion showed that the intrinsic geometrical properties of the liquid film and `droplet' are extremely important. We shall also see that a number of other length scales emerge from the problem in due course. However, we shall use the length scales already seen to non-dimensionalize the equilibrium problem: we use $h_0$ and $x_c$ to non-dimensionalize the  vertical and horizontal directions, respectively, giving the natural dimensionless variables
\begin{equation}
     X=x/x_c,\quad H=h/h_0, \quad\text{and}\quad \Pi(X)=\frac{x_c^2}{\gamma h_0}p_e(x).
     \label{eqn:StaticND}
\end{equation}
(Note that the scaling analysis suggests that the natural scale for $h_c=h(x_c)$ is $x_c$; we use $h_0$ to rescale $h(x)$ to capture the natural behaviour far from the triple line.) Substituting these relationships into \eqref{eq:static}, we find that
\begin{equation}
    \Pi(X)- \frac{H_{XX}}{\left(1+H_X^2/\As^2\right)^{3/2}} - \frac{\AvdW}{H^{3}} +\Bo\, H - P_\infty =0,
    \label{eq:Press_rescaled}
\end{equation}
for $0\le X \le \Xinf=\xinf/x_c$, where 
\begin{equation}
    \As=x_c/h_0\gg1,\quad   \Bo=\frac{\rho g  x_c^2}{\gamma}
    \label{eqn:AspectRatioDefn}
\end{equation} 
are the aspect ratio of the film  and the Bond number of the `droplet', respectively, while
\begin{equation}
    \AvdW=\frac{A x_c^2}{\gamma h_0^4}=\frac{A\As^4}{\gamma x_c^2}
    \label{eqn:vdWNumberDefn}
\end{equation} measures the relative importance of van der Waals forces ($A/h_0^3$) and the typical capillary  pressure  in the original lubricating film ($\gamma h_0/x_c^2$) and
$P_\infty=p_\infty x_c^2/(\gamma h_0)$. 

We have already argued that the hydrostatic pressure term must play an important role in determining the equilibrium of the meniscus beyond the triple line (see also Appendix \ref{sec:Experiments}). It is therefore tempting to ignore the role of van der Waals forces entirely, especially since the dimensionless parameter $\AvdW\ll1$. However, $\AvdW$ measures the importance of van der Waals pressures for a film thickness on the order of the initial film thickness: in equilibrium,  the meniscus beneath the droplet has typical thickness \cite{Daniel2017} $h\approx\heqm= (A x_c/\gamma)^{1/3}\approx 20 \mathrm{~nm} \ll h_0$ significantly amplifying the importance of the term $\AvdW/H^3$ in \eqref{eq:Press_rescaled}.

With the above non-dimensionalization, the boundary conditions for the solution of \eqref{eq:Press_rescaled}  become 
\begin{equation}
    H_X(0)=H_X(\Xinf)=0\quad\text{and}\quad\int_0^{\Xinf} H\,\dd X = \Xinf
    \label{eq:BC_static}
\end{equation} while the jump conditions across the triple line \eqref{eq:staticsBCs2} become
\begin{equation}
    H(1^+)-H(1^-)=0\quad \text{and}\quad \left. \frac{H_X}{\left(\As^2+H_X^2\right)^{1/2}}\right\vert_{1^+}-\left. \frac{H_X}{\left(\As^2+H_X^2\right)^{1/2}}\right\vert_{1^-}=-\cos{\dt},
    \label{eq:matchingcondt}
\end{equation}
with the rotation angle $\dt$  satisfying  the dimensionless version of \eqref{eqn:CLAsymmetry}, namely
\begin{equation}
    \dt=\sin^{-1}\left[\left.{\left(1+H_X^2/\As^2\right)^{-1/2}}\right\vert_{1^+}-\left.{\left(1+H_X^2/\As^2\right)^{-1/2}}\right\vert_{1^-}\right].
    \label{eq:RotAng}
\end{equation}

\subsection{Numerical results\label{sec:NumsMainText}}

\begin{table*}[!]
\caption{\label{tab:table1} {A summary of the typical values of control parameters with  references to experimental work} (upper part of the table), and the corresponding emergent dimensionless parameters and time scales (lower part of the table).}
\begin{ruledtabular}
\begin{tabular}{c|ccccc}
&$x_c$&$\gamma$&$A$&$h_0$&$\mu$\\
Value&$\approx0.5\mathrm{~mm}$&$10-60 \mathrm{~mN/m}$&$\sim10^{-21}\mathrm{~J}$&$2-20\mathrm{~\mu m}$&$\sim10\mathrm{~mPa\,s}$\\
Reference&\cite{Schellenberger2015,Daniel2017}&\cite{Kreder2018,Mchale2019,Schellenberger2015}&\cite{Schellenberger2015,Daniel2017}&\cite{Daniel2017,Kreder2018,Mchale2019}&\cite{Kreder2018,Daniel2017}\\\hline\hline
&$\As$&$\Bo$&$\AvdW$&$\tau_\ast$&$\tau$\\
Value&$25-250$&$10^{-2}-10^{-1}$&$10^{-7}-10^{-3}$&$2-20\mathrm{~\mu s}$&$10-10^4\mathrm{~s}$\\
\end{tabular}
\end{ruledtabular}
\vspace{0.5cm}
\end{table*}

We solve the problem \eqref{eq:Press_rescaled} subject to \eqref{eq:BC_static}--\eqref{eq:matchingcondt} numerically on the interval $0\leq X\leq \Xinf$  using the multi-point boundary value problem feature of \texttt{bvp5c} in \textsc{MATLAB}. Typical values of the control parameters of the system are summarized in Table \ref{tab:table1}. In many applications it is droplets of millimetric diameter that are of most interest, so that $x_c= 0.5\mathrm{~mm}$ is typical. We then have $\Bo\approx 10^{-2}$. The far-field thickness of the oil layer may vary significantly depending on surface preparation, but is typically on the order of $h_0=5\mathrm{~\mu m}$, corresponding to $\As\approx 10^2$. The typical value of the Hamaker constant is $A=10^{-21}\mathrm{~J}$ so that  $\AvdW\approx\As^4/10^{12}$. (Note that when the drop size is fixed through a constant $\Bo$, increasing the aspect ratio $\As$ corresponds to decreasing the initial film thickness provided that $\AvdW$ is chosen such that $\AvdW\propto\As^4$.)

\begin{figure}[!h]
    \centering
    \includegraphics[width=14.5cm]{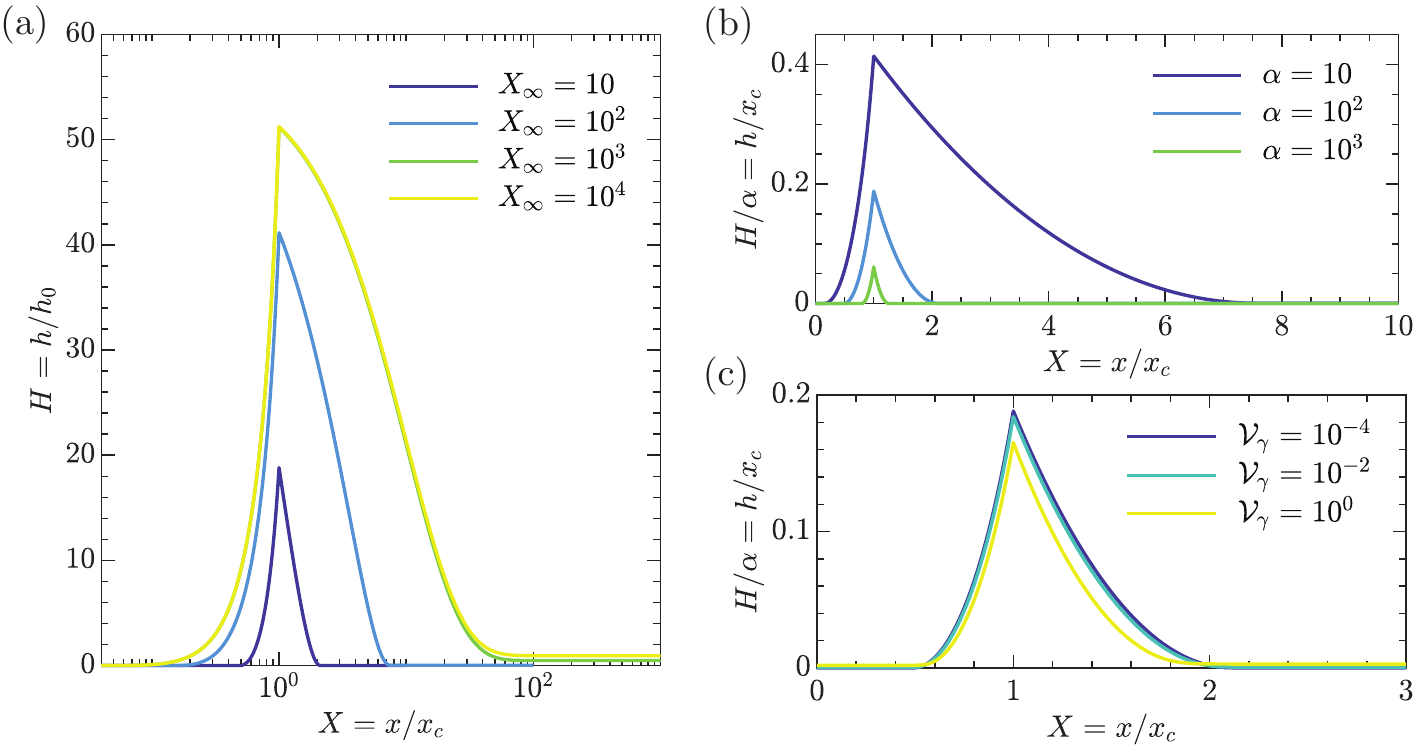}
    \caption{Equilibrium interface profiles calculated numerically from the solution of \eqref{eq:Press_rescaled}. (a) The effect of system size, $\Xinf=x_\infty/x_c$, with  $\Bo=10^{-2},\AvdW= 10^{-4}$, $\As=10^2$ fixed. (b) The effect of initial film thicknesses (for fixed plate size $\Xinf=10$ and Bond number, $\Bo=10^{-2}$) is obtained by changing $\As^{-1}\in [10^{-3},10^{-1}]$ but maintaining $\AvdW={\As}^4/10^{12}$. (a) and (b) show that equilibria are significantly affected by the size of the reservoir of oil available to be sucked into the skirt. (c) The effect of van der Waals forces as encoded by the value of $\AvdW$ is much more limited, as shown by profiles calculated with fixed $\Bo=10^{-2}$, initial film thicknesses (so that $\As=10^{2}$) and plate size $\Xinf=10$. }    \vspace{0.5cm}
    \label{fig:StaticProfile}
\end{figure}

\begin{figure}
    \centering
    \includegraphics[width=14.5cm]{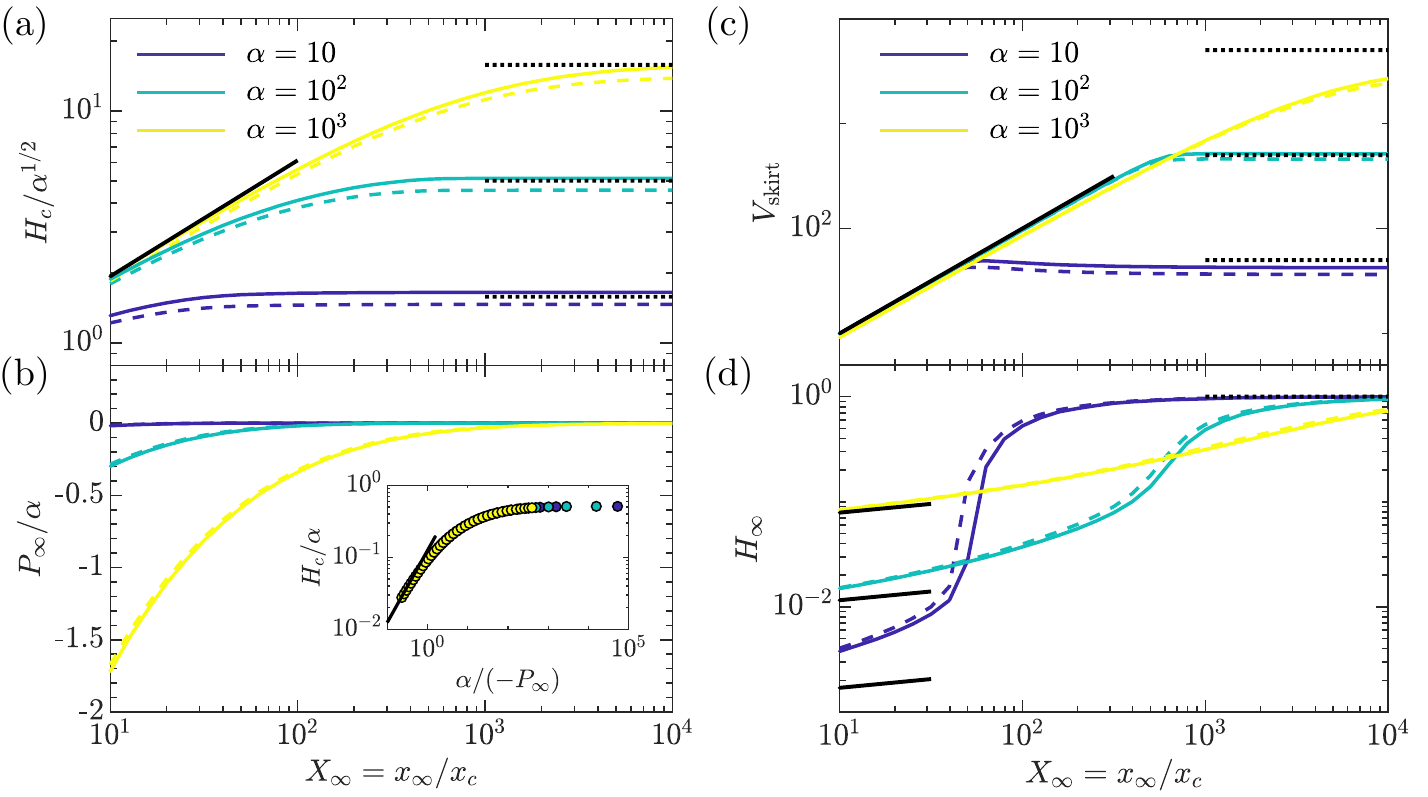}
    \caption{Dependence of (a) the equilibrium skirt height at the triple line, $H_c$, (b) the pressure in the film, $\Pinf$, (c) the skirt volume, defined by Eq.~\eqref{eq:SkrVol}, and (d) the far-field equilibrium film thickness, $\Hinf$, as functions of the relative plate size, $X_\infty$. Asymptotic results, derived in \S\ref{sec:AsyStatics}, for lubricant-sated systems (dotted black lines) and lubricant-starved systems (solid black lines) are also shown, given by \eqref{eqn:LargeSystemHinfandV}, \eqref{eq:largeskr} and \eqref{eq:SmallSkirtAsy}. The solid line in the inset of (b) is based on $H_c=\As^2/(-8\Pinf)$, which is given by combining \eqref{eq:SmlSkirtPres} and \eqref{eq:SmallSkirtAsy}. All calculations use $\Bo=10^{-2}$ with $\AvdW=\As^4/10^{12}$ and $\As$ varying from $10$ to $10^3$. All solid and dashed curves of the same colour are calculated using the same parameters but with the dashed curves using the small slope approximation of the curvature, $|h_x|\ll1$, and the no-rotation assumption for the pulling force, i.e.~$\Delta\theta=0$. }    \vspace{0.5cm}
    \label{fig:HeightnVolume}
\end{figure}

Results for typical parameter values are shown in fig.~\ref{fig:StaticProfile} and fig.~\ref{fig:HeightnVolume}. For comparison, we also solve the linearized problem, $|H_X|\ll\As,\dt\ll1$ obtaining qualitatively similar results --- compare the dashed and solid curves in fig.~\ref{fig:HeightnVolume} --- even though the angles near the triple line are \emph{not} small.

In these simulations, we fix the size of the droplet with $\Bo=10^{-2}$  and  vary $\As$ and $\Xinf=x_\infty/x_c$ to study the role played by the initial film thickness and the size of the supporting plate on the final skirt shape.  Some examples of these skirt profiles are shown in  fig.~\ref{fig:StaticProfile}a,b and demonstrate that the size of the system ($\Xinf$) and the film thickness (through $\As$) both play important roles in determining the final equilibrium skirt shape and volume, as might have been expected from the earlier scaling discussion. However, the role of van der Waals forces in  the final interface shape at a macroscopic scale is minimal unless $\AvdW$ approaches a much larger value (see fig.~\ref{fig:StaticProfile}c, also discussed in Appendix \ref{sec:VgEffectStatics}). (Note that the typical value of $\AvdW$ given in Table~\ref{tab:table1} is calculated based on oil layer thicknesses ranging from 2 to 10 $\mathrm{\mu m}$ \cite{Daniel2017,Kreder2018,Mchale2019}; however, it may be possible to reduce the layer thickness, and hence achieve a large $\AvdW$, by using volatile lubricant, as observed in one unusual drop-lubricant system \cite{Tress2017}.)

Beyond the skirt profile itself, the key features of the skirt are its height, the liquid pressure within it, the volume of liquid captured within it, and the asymmetry in the meniscus shape about the triple line. The dependence of the skirt height 
\begin{equation}
    H_c=H(X=1)
\end{equation} on the initial film thickness and the size of the plate is shown in fig.~\ref{fig:HeightnVolume}a.

Similarly, the predicted pressure within the liquid skirt, $\Pinf$, is shown in fig.~\ref{fig:HeightnVolume}b as a function of system size, $\Xinf/X_c$, for different film thicknesses. Previously, \citet{Semprebon2017,Semprebon2021} obtained a relationship between the triple line height, $H_c$, and the radius of curvature of the meniscus, $\kappa\propto \Pinf$, under the assumption that the skirt remains small compared to the droplet.
Such geometrical arguments are entirely compatible with our approach since the film pressure is determined as part of the solution; indeed, plotting $H_c/\As=h_c/x_c$ as a function of $\As/(-\Pinf)$ produces a collapse of our numerical data onto a master curve that is very similar to that provided by \citet{Semprebon2021} (compare the inset of fig.~\ref{fig:HeightnVolume}b with fig.~4 of \citet{Semprebon2021}). However, we emphasize that our result goes beyond that of \citet{Semprebon2021} by determining the height of the triple line and the curvature/pressure of the skirt  in terms of the key control parameters within the system, namely $\Xinf$ and $\As$. In particular, we will determine  explicit relationships for $H_c$ and $\Pinf$ as functions of the system size, showing that the skirt's equilibrium may be `lubricant-sated' or `lubricant-starved' depending on how much lubricant is available to the skirt, as will be discussed in sections \ref{sec:LargeSystems} and \ref{sec:SmallSystems}, respectively.

A key question is then whether the liquid in the skirt comes predominantly from the liquid beneath the droplet being squeezed out into the skirt, or rather is sucked into the skirt from the remainder of the lubricating layer. To  answer this, we define the skirt volume by the volume of liquid lifted above the far-field liquid level, i.e. 
\begin{equation}
    \Vskirt=\int_{X^-}^{X^+}\left[H(X)-H(\Xinf)\right]\,\dd X,
    \label{eq:SkrVol}
\end{equation} where the limits of integration $X^\pm$ are identified such that the interval $(X_-,X_+)$ is the largest interval containing the triple line $X=1$ throughout which $H\geq 1.01 H(\Xinf)$ (i.e.~$X^-<1<X^+$, and $H(X^\pm)=1.01H(\Xinf)$ with $H(X)\geq1.01H(\Xinf)$ for all $X\in[X_-,X_+]$). With this definition, a value of $\Vskirt\approx1$ suggests that the skirt is dominated by liquid that is pushed into it from beneath the droplet (since the liquid layer beneath the droplet in equilibrium $\Heqm=(\AvdW/\As)^{1/3}\sim10^{-2}\ll1$, only a negligible amount of the film is expected to remain there and a dimensionless volume of liquid $\approx1$ is pushed out from under the droplet). Conversely, a value $\Vskirt\gg1$ suggests instead that the skirt is dominated by liquid that is sucked into it from the remainder of the lubricant film. The dependence  of the skirt volume defined in this way on the system size, $\Xinf$, is shown in fig.~\ref{fig:HeightnVolume}c. Crucially, we see that $\Vskirt\gg1$ for all but the very smallest system sizes, indicating that the majority of the liquid within the skirt has been sucked into it from the reservoir, rather than being squeezed out from beneath the `drop'.
 
Figure~\ref{fig:HeightnVolume}d shows the dependence of the equilibrium far-field film thickness on  the system size, $\Xinf$. As might be expected, this shows a significant change in thickness compared to the initial condition when the system is small, while for large systems the far-field thickness is essentially unchanged from its initial value.

 The results in fig.~\ref{fig:HeightnVolume} show that, unless the system is extremely large, the size of the skirt depends sensitively on how large the system is, $X_\infty$.  The dependence on $\Xinf$ arises from the global conservation of mass: the skirt requires liquid to be supplied by the coating of the remainder of the plate (since $\Vskirt\gg1$, as already discussed), and hence on how much liquid is available on the plate. We also see that what a `large' system size means depends on the thickness of the liquid film: for smaller $\As$ (i.e.~thicker films with drop size fixed) the skirt volume saturates at smaller system sizes, again because more liquid is available to the skirt. We therefore refer to such systems as lubricant-sated, because the skirt's appetite for lubricating fluid is satisfied, rather than talking about the system size. 
 
 \begin{figure}[!h]
    \centering
    \includegraphics[width=9cm]{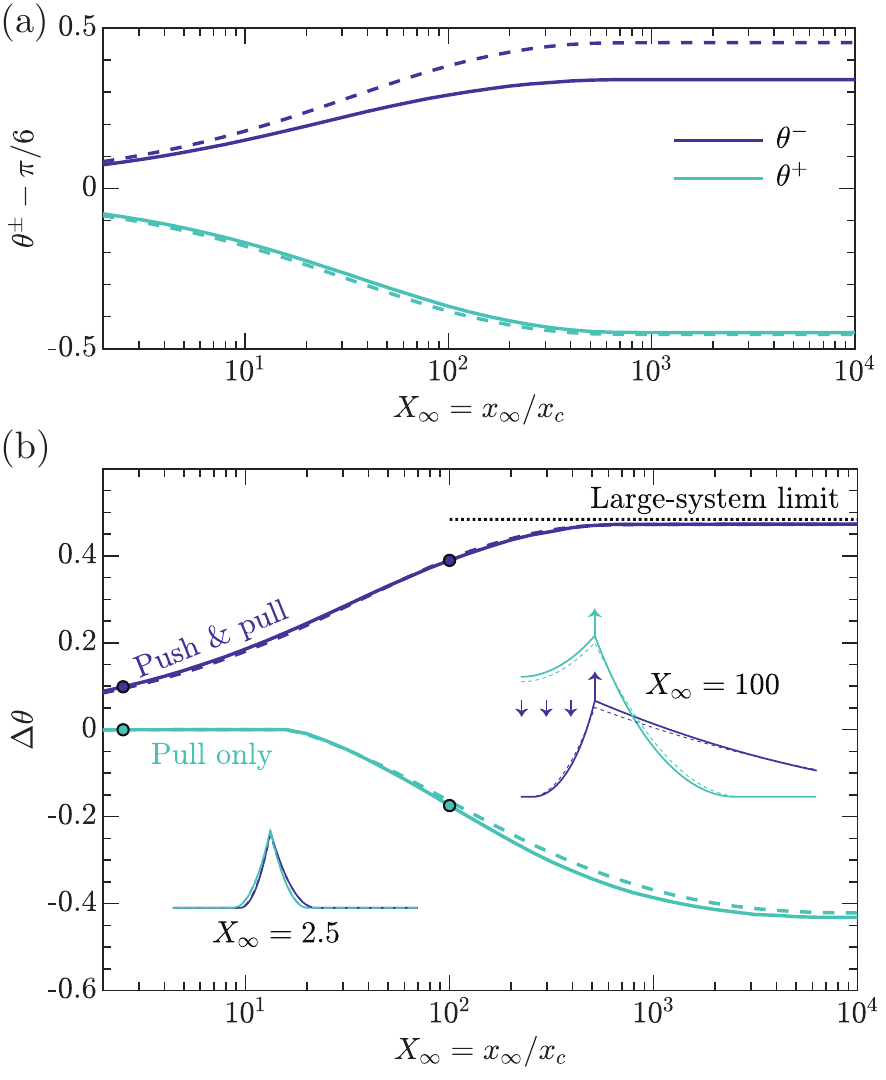}
    \caption{(a) Deviation of the meniscus angles at the triple line (as defined in fig.~\ref{fig:Schematic}) from $\pi/6$ as a function of the system size, $\Xinf$. (b) Equilibrium values of the rotation angle $\Delta\theta$ defined by Eq.~\eqref{eq:RotAng} for the scenarios in which there are only meniscus forces (`pull only') and that in which the droplet's capillary pressure also pushes the interior meniscus downward (`push and pull'). Note that the `self-interaction' of the droplet's two menisci causes a clockwise rotation of the menisci $\dt<0$, while the effect of the droplet's positive Laplace pressure works in the opposite direction, causing an anti-clockwise rotation, $\dt>0$.}    \vspace{0.5cm}
    \label{fig:Cheerios}
\end{figure}

 The final quantity of interest in the static problem is the asymmetry of the skirt about the triple line, which is defined as $\theta^+-\theta^-$. This asymmetry is intimately related to the rotation of the vertical pulling force, $\dt$, since the triple line must remain in vertical and horizontal force equilibrium. (In particular, for perfect symmetry $\theta^\pm=\pi/6$ and $\dt=0$ to ensure all included angles are $2\pi/3$.)  This asymmetry can be seen in fig.~\ref{fig:Cheerios}a where the deviation of the local  angle from $\pi/6$  on the left side (inner meniscus) and right side (outer meniscus) are shown as a function of system size; crucially the deviation from symmetry increases with the system size ($\Xinf$). 
 
 The rotation of the pulling force, $\dt$, is shown in fig.~\ref{fig:Cheerios}, and demonstrates that the Neumann triangle generically rotates anti-clockwise (since $\dt>0$); $\dt$ also increases with system size (as might be expected since meniscus symmetry and rotation are intricately related). On the face of it there are two possible sources for this rotation: the applied squeezing pressure pushes the inner and outer portions of the meniscus differently and hence may lead to rotation while the  interaction between the two menisci at $\pm X_c$ also leads to an asymmetry, and hence rotation. This latter effect is somewhat analogous to the `Cheerios effect' \cite{Vella05}, and can be isolated from the former effect in our model by omitting the indicator function term in \eqref{eq:PressDis}; fig.~\ref{fig:Cheerios} therefore shows the rotation angle $\dt$ predicted from this model in which the interface is subject only to a `pull' from surface tension, without the `push' from the capillary pressure. Interestingly, the rotation with only the `pull' effect is in the opposite sense to that obtained from the full model; we conclude that the interaction between the menisci is not the dominant source of this asymmetry --- rather it is the effect of the droplet's squeezing pressure that is  crucial.

Note that in all cases our numerical results with the fully nonlinear curvature and rotation $\dt$ are only slightly different to these with linearized curvatures and no rotation --- compare dashed and solid curves in figs.~\ref{fig:HeightnVolume} and \ref{fig:Cheerios}. However, this does \underline{\underline{not}} indicate that slopes remain small; rather there seems to be a fortuitous cancellation of the two effects.

\section{Asymptotic results for equilibrium \label{sec:AsyStatics}}

Our numerical results have revealed quite different behaviours depending on the amount of lubricant available in the system relative to that needed for a full skirt, $\xinf h_0/x_c^2=\Xinf/\As$: for small $\Xinf$ and large $\As$, the availability of lubricant is the limiting factor in the skirt's growth; we therefore refer to such systems as \emph{lubricant-starved} (following \citet{Tress2017}). For such systems, the menisci are approximately symmetrical, with the droplet's line force acting approximately vertically; for large $\Xinf/\As$ (i.e.~lubricant-sated systems), the outer meniscus is essentially horizontal with the droplet's two menisci inclined at $\pi/6$ to the horizontal to balance one another. To understand the behaviors in these two limits further, we turn now to consider the limits of lubricant-sated and starved systems in turn.

\subsection{Lubricant-sated systems \label{sec:LargeSystems}}

The numerical results of \S\ref{sec:NumsMainText} suggest that when $\Xinf/\As$ is sufficiently large, the system approaches a well-defined limit in which further increases in $\Xinf$ do not change the properties of the final equilibrium: there is enough liquid coating the substrate for the skirt to find its preferred equilibrium. In this lubricant-sated limit there is a small, but positive, pressure throughout the film since $\Pinf=\Bo\,\Hinf-\AvdW/\Hinf^3$, $\AvdW\ll\Bo\ll1$ in general and the far-field film thickness is close to its initial value, i.e.~$\Hinf \sim 1$. In this case the two menisci that meet at the triple line behave very differently: the outer meniscus is affected only by gravity (since $\AvdW\ll1$) and so decays according to the usual balance between hydrostatic pressure and capillarity to give 
\begin{equation}
    H \approx (H_c-\Hinf) \exp{\left[-\sqrt{\Bo}(X-1)\right]}+\Hinf,
    \label{eq:outerprofile}
\end{equation}
for $X>1$, where we have assumed $|H_X|\ll\As$ (i.e.~small slope deformations), which is self-consistent since we expect the constant $H_c\sim O(\As)$ and then \eqref{eq:outerprofile} gives that $|H_X|\sim \As\sqrt{\Bo}\ll\As$.  Equation \eqref{eq:outerprofile} also allows us to calculate that
\begin{equation}
    \Hinf= 1+O\left(\frac{H_c}{\Xinf\Bo^{1/2}}\right)\quad\text{and}\quad\Vskirt =\frac{H_c}{\sqrt{\Bo}}\left[1+O(H_c^{-1})\right].
    \label{eqn:LargeSystemHinfandV}
\end{equation}

To solve the shape of the inner meniscus, we neglect the $\AvdW/H^3$ term (since it decays quickly as the film height increases beyond $\Heqm\ll 1\ll H_c$) and integrate \eqref{eq:Press_rescaled} once to have
\begin{equation}
    \left(1+\tfrac{1}{\As^2}H_X^2\right)^{-1/2} + \tfrac{1}{2\As^2}\Bo\,H^2 + \tfrac{1}{\As}H\cos\dt - 1=0,
    \label{eq:innerprofile}
\end{equation}
for $0<X<1$, where we have assumed $H_X=0$ as $H\to0$ beneath the drop to determine the constant of integration. To solve for $H_c$ and $\dt$, instead of using the vertical and horizontal force balance conditions in \eqref{eq:matchingcondt} and \eqref{eq:RotAng}, we exploit their geometrical counterparts: $\theta^+ + \dt =\pi/6$ and $\theta^+ + \theta^- = \pi/3$, where $\theta^+\approx |H_X(1^+)/\alpha|$ can be calculated using \eqref{eq:outerprofile} while $\theta^-=\tan^{-1}\left[{H_X(1^-)/\alpha}\right]$ can be given by \eqref{eq:innerprofile}. Letting $H_c=\As(\eta_0+\delta\eta_1)$ and $\dt=\pi/6+\delta\dt$ and $\Bo\ll1$ we obtain:
\begin{equation}
    H_c\approx\frac{\As}{3\sqrt{3}}(3-4\Bo^{1/2})\quad\text{and}\quad\dt \approx \tfrac{1}{6} \pi -\tfrac{1}{\sqrt{3}} \Bo^{1/2},
    \label{eq:largeskr}
\end{equation}
where we assumed $H_c\gg 1$ for the calculation of $H_c$. Equations \eqref{eqn:LargeSystemHinfandV} and \eqref{eq:largeskr} are used to plot the dotted lines in fig.~\ref{fig:HeightnVolume}a and fig. \ref{fig:Cheerios}b, and agrees well with our numerical results in the appropriate limit. Moreover, the size of the error term in the first part of \eqref{eqn:LargeSystemHinfandV} explains why the lubricant-sated regime is not reached unless $\As/(\Xinf\Bo^{1/2})\ll1$.

\subsection{Lubricant-starved systems\label{sec:SmallSystems}}

When $\As$ is large but $\Xinf$ is not sufficiently large to make $\As/(\Xinf\Bo^{1/2})\ll1$, our numerical results show that the skirt region is small and the far-field film thickness $\Hinf\ll1$: the vast majority of liquid initially coating the substrate is drawn into the skirt by the skirt's negative capillary pressure. As such, the skirt's growth is limited by the amount of liquid available to it. Moreover, the pressure within the skirt is relatively large in magnitude. 

In this volume-limited limit, we expect  the height and radius of curvature of the skirt to be comparable and hence the meniscus height not to be large enough for a significant deviation from the ideal Neumann triangle to take place: we expect the angles at the triple line to be close to $\pi/6$, with relatively little meniscus rotation, $\dt\ll1$. These expectations are borne out by our numerical results. To understand this limit further, we follow a similar approach to that used in \cite{Semprebon2017,Kreder2018}, approximating the meniscus shape on both sides of the triple line by a parabola with a radius of curvature much smaller than  unity (i.e.~the skirt is much smaller than the `drop'). This small-skirt assumption gives symmetric meniscus profiles about the triple line, since there is relatively little meniscus rotation (fig.~\ref{fig:Cheerios}b), and so we may write:
\begin{equation}
    H\approx-\frac{1}{2}P_\infty\left(X_s-|X-1|\right)^2
\end{equation}
for $1-X_s\le X \le 1+X_s$, with $2X_s\ll1$ the horizontal extent of the skirt. The slope discontinuity condition, \eqref{eq:matchingcondt}, and volume-conservation conditions, \eqref{eq:BC_static}, can be used to solve for $X_s$ \text{and} $P_\infty$. In particular, we find
$\Pinf=-\As/(2X_s)$ and $\Xinf=-\Pinf X_s^3/3$ so that
 \begin{equation}
    \Pinf=-\left(\frac{\As^3}{24\Xinf}\right)^{1/2},
    \label{eq:SmlSkirtPres}
\end{equation} which immediately leads to 
\begin{equation}
    H_c=\sqrt{\frac{3}{8}}\As^{1/2}\Xinf^{1/2},\quad \Hinf=\left(\frac{24\AvdW^2\Xinf}{\As^3}\right)^{1/6},\quad \Vskirt = \Xinf ,\quad\text{and}\quad \dt = 0.
    \label{eq:SmallSkirtAsy}
\end{equation}
Here the far-field thickness is determined from the stabilization of the film by van der Waals pressure, i.e.~$\Hinf\approx(-\AvdW/\Pinf)^{1/3}$. It is also worth noting that the  largest unscaled slope of the meniscus here is $|h_x|=\As^{-1}H_X(1)=\As^{-1}\Pinf X_s=1/2$, which is an $O(1)$ quantity: again, even though the deformations in this case are relatively small, the interfacial slopes may be $O(1)$.

The first two results in \eqref{eq:SmallSkirtAsy} are plotted as the solid curves in fig.~\ref{fig:HeightnVolume}. We find good agreement with numerical results for small skirts as these expressions are derived based on the assumption that $X_s\ll 1$. (Note also that the prediction $\Vskirt=X_\infty$ is merely a consistency check since we  assumed in the above derivation that the growth of the skirt is limited by the volume of liquid available.) Finally, we note that the small-skirt assumption $X_s\ll 1$ requires the control parameters of the system to satisfy $\Xinf/\As\ll 1$, i.e.~$\xinf/x_c\ll x_c/(6h_0)$ in dimensional form, just as we had discussed based on intuitive grounds already. 

\subsection{Transition from lubricant-starved to lubricant-sated systems\label{sec:Transition}}

In the preceding sections, we have calculated the dependence of the final skirt volume on the volume of  lubricant available. The numerical results in fig.~\ref{fig:HeightnVolume}c show that, on the whole, the transition between these relations is relatively sudden, being where the asymptotic results for the skirt volume given in \eqref{eq:SmallSkirtAsy} and \eqref{eqn:LargeSystemHinfandV} intersect. This transition happens when
\begin{equation}
    \xinf\sim \frac{x_c}{h_0}\ell_c
    \label{eqn:largeSystemsDefn}
\end{equation} 
or 
\begin{equation}
    X_\infty\sim \As\,\Bo^{-1/2}.
\end{equation} 
Physically, this result suggests that large systems are those for which the total amount of liquid coating the substrate, $\xinf\times h_0$, is significantly larger than the volume that is required to make a static meniscus of height $\sim x_c$, making use of \eqref{eq:largeskr} and width $\ell_c$.  

It is also worth noting that the system size at which the skirt becomes `sated' (and hence no longer limited by the volume of liquid in the coating) is the capillary length multiplied by a factor $x_c/h_0$; for a  droplet with the parameters suggested in table \ref{tab:table1}, this dimensionless factor is on the order of $100$, so that experiments may well not be in the lubricant-sated limit unless the system size reaches tens of centimeters (assuming $\lc\approx 2\mathrm{~mm}$). As such, the skirts observed experimentally are likely to be at least partially limited by the volume of liquid available within the lubricant layer.

The limiting effect of the available lubricant would be somewhat reduced in axisymmetry, rather than the two-dimensional problem considered here, since then the volume of lubricant available increases quadratically with the plate size (rather than linearly). Following the idea that led to \eqref{eqn:largeSystemsDefn}, the transition from small to large systems in an axisymmetric system is expected to occur when the volume of the lubricant, $\xinf^2\times h_0$, becomes comparable to the volume required to make a skirt of height $\propto x_c$ and projected area $(x_c+\lc)^2\approx\lc^2$, i.e.~$\xinf\propto\alpha^{1/2}\lc$. Therefore, a centimeter-sized plate with a single deposited droplet  may be close to the lubricant-sated limit, but would not be with multiple drops, for example. This immediately raises the question of why so many small skirts (compared to the capillary length) have been observed in previous experiments, see for example \cite{Kreder2018,Daniel2017,Keiser2020,Mchale2019}. We therefore turn to study the dynamic process through which equilibrium is established, seeking to understand the time scale on which this equilibrium is reached. We shall see that though van der Waals forces play little role in the macroscopic properties of the final equilibrium, they \emph{are} important in determining the time taken to setup the equilibrium, which, as a result, can be extremely long.

\section{Dynamics: Formulation and early-time behavior \label{sec:Dynamics}}

\subsection{The dynamic push-and-pull model}

A natural choice to model the slow motion of lubricant in response to the pressure field \eqref{eq:PressDis} is the long-wavelength approximation of the Stokes equations (cf.~lubrication theory) \cite{Leal2007}. Given that the behavior near the triple line requires the interface slope to be $O(1)$, i.e.~not small, it is natural to wonder whether this long wavelength approximation is valid. It is therefore worth noting that modifications to lubrication theory to account for such slopes have been proposed previously \cite{Snoeijer2006,Tavakol2017}, while the differences with standard lubrication theory are generally small and quantitative, rather than qualitative. In addition, we have already seen in the equilibrium problem that the fully nonlinear and linear problems agree very well, modulo some small quantitative differences. We shall therefore use this linearization  (i.e.~we use the linearized curvature and neglect the rotation of the pulling force) throughout our study of the dynamics of the problem, combined with the long-wavelength approximation of the Stokes equations, or lubrication theory. 

A standard analysis \cite{Leal2007} shows that the evolution of the film thickness $h(x,t)$ is described by Reynolds' equation:
\begin{equation}
    \frac{\partial h}{\partial t} = \frac{1}{3\mu}\frac{\partial}{\partial x}\left(h^3\frac{\partial p}{\partial x}\right),
    \label{eq:Reynolds}
\end{equation}
where $\mu$ is the viscosity of the liquid in the film. Note that in deriving \eqref{eq:Reynolds} we have used boundary conditions of no-slip on the solid surface ($z=0$) and no shear stress at the oil-drop interface (since the droplet is usually significantly  less  viscous than  the  oil  layer); if both boundaries were  no-slip, the factor $3$ in \eqref{eq:Reynolds} would instead be $12$.

Substituting the pressure field from \eqref{eq:Pressure} into \eqref{eq:Reynolds} assuming small slopes within the film ($|h_x|\ll1$) for consistency with the derivation of \eqref{eq:Reynolds} and neglecting the rotation of the pulling force, we obtain a nonlinear diffusion equation for the film thickness
\begin{equation}
    \frac{\partial h}{\partial t} = \frac{1}{3\mu}\frac{\partial}{\partial x}\left[h^3\left(\frac{\partial p_e}{\partial x}-\gamma h_{xxx}+\frac{3A}{h^4}h_x+\rho g h_x\right)\right],
    \label{eq:EvlEq}
\end{equation}
for $t>0$. 

Equation \eqref{eq:EvlEq} requires an initial condition and four boundary conditions. We shall use a uniform initial profile of the film:
\begin{equation}
    h(x,0)=h_0
    \label{eqn:ICs}
\end{equation} for simplicity. We shall also assume that the problem remains symmetric about $x=0$; we therefore consider only $0\le x \le \xinf$ and have immediately  symmetry conditions at $x=0$, combined with requirements that  at the edge of the plate, $x_\infty$, the film slope and fluid flux should vanish i.e.
\begin{equation}
    h_x(0,t)=h_x(x_\infty,t)=h_{xxx}(0,t)=h_{xxx}(x_\infty,t)=0.
    \label{eqn:BCs}
\end{equation} 
While we shall predominantly solve the above problem numerically, the use of linearization (i.e.~assuming small slopes and neglecting the rotation of the pulling force) allows some analytical descriptions of the  evolution of the initially uniform film at early times.

\subsection{Early-time behaviour}
    
At early times, the film beneath the `droplet' does not yet `know' the final equilibrium thickness that it will reach. As a result, the relevant vertical length scale with which to measure deformations of the thin liquid film is $h_0$. Similarly, we shall find that the early motion is dominated by the pulling force at the triple line, which does not depend on the droplet size, so that the relevant horizontal length scale is also $h_0$. A scaling analysis of the dimensional governing equation \eqref{eq:EvlEq} suggests that the relevant time scale for these early dynamics is
\begin{equation}
    \tau_*=\frac{3\mu h_0}{\gamma}.
    \label{eqn:EarlyTimescale}
\end{equation}
Typically,  $\tau_\ast$ is on the order of $1\mathrm{~\mu s}$ (see Table \ref{tab:table1}).

\subsubsection{Non-dimensionalization}

 Based on the previous discussion, we introduce a slightly different  non-dimensionalization for length scales at early times to that used for the  static model, namely we let
\begin{equation}
    \eX=x/h_0   ,\quad \eH=h/h_0, \quad\text{and}\quad \eT=t/\tau_*,
\end{equation} so that the governing equation \eqref{eq:EvlEq} becomes
\begin{equation}
    \frac{\partial \eH}{\partial \eT} = \frac{\partial}{\partial \eX}\left[\eH^3\left(\frac{\partial\ePi}{\partial \eX} -\frac{\partial^3 \eH}{\partial \eX^3}\right)+\frac{1}{\As^2}\left(\frac{3\AvdW}{\eH}+\Bo \eH^3\right)\frac{\partial \eH}{\partial \eX}\right], 
    \label{eq:EvlEq_Early}    
\end{equation}
for $0\le \eX \le \As\Xinf$, and where $\ePi(\eX)=\ind(\eX)/\As-\delta(\eX-\As)$ for $\eX > 0$.  Note that \eqref{eq:EvlEq_Early} involves the already familiar parameters, e.g.~$\As$, $\Bo$, and $\AvdW$ that were defined in \eqref{eqn:AspectRatioDefn} and \eqref{eqn:vdWNumberDefn}.

With this rescaling, the  initial and boundary conditions become
\begin{equation}
    \eH(\eX,0)=1,
    \label{eq:IntCd_Early}
\end{equation} and
\begin{equation}
    \left.\frac{\partial \eH}{\partial \eX}\right\vert_{(0,\eT)}=\left.\frac{\partial \eH}{\partial \eX}\right\vert_{(\As\Xinf,\eT)}=\left.\frac{\partial^3 \eH}{\partial \eX^3} \right\vert_{(0,\eT)}=\left.\frac{\partial^3 \eH}{\partial \eX^3} \right\vert_{(\As\Xinf,\eT)}=0.
    \label{eq:BC_Early}
\end{equation}

The problem specified in \eqref{eq:EvlEq_Early}--\eqref{eq:BC_Early} is solved  numerically using the method of lines implemented in MATLAB, following \citet{shampine2007}. More details of our numerical method are given in Appendix \ref{sec:Numerics}. However, we note here that a discontinuity in the driving pressure, $\ePi(\eX)$,  is introduced by the indicator function and Dirac-$\delta$ function at the triple line (which is now located at $\eX=\As$). This is treated by considering the regions with $\eX<\As$ and $\eX>\As$  separately and connecting them via four matching conditions at the triple line. Specifically, the film thickness is continuous, so that
\begin{equation}
    [\eH]_-^+=0,
    \label{eq:matching1}
\end{equation}
with $[g]_-^+=g(\As^+)-g(\As^-)$, while there are jumps in the slope and curvature that can be written
\begin{equation}
    \left[\frac{\partial \eH}{\partial \eX}\right]_-^+=-1,\quad\text{and}\quad\left[\frac{\partial^2 \eH}{\partial \eX^2}\right]_-^+=-1/\As.
\end{equation} A final jump condition is provided by the conservation of fluid flux across the triple line; this is simplified to
\begin{equation}
    \left[\frac{\partial^3 \eH}{\partial \eX^3}\right]_-^+=0,
    \label{eq:matching4}
\end{equation}
by neglecting the van der Waals and hydrostatic pressure terms, which would provide a correction of order $\sim({3\AvdW}+\Bo)/\As^2 \lll 1$. 

\begin{figure}[ht]
    \centering
    \includegraphics[width=10cm]{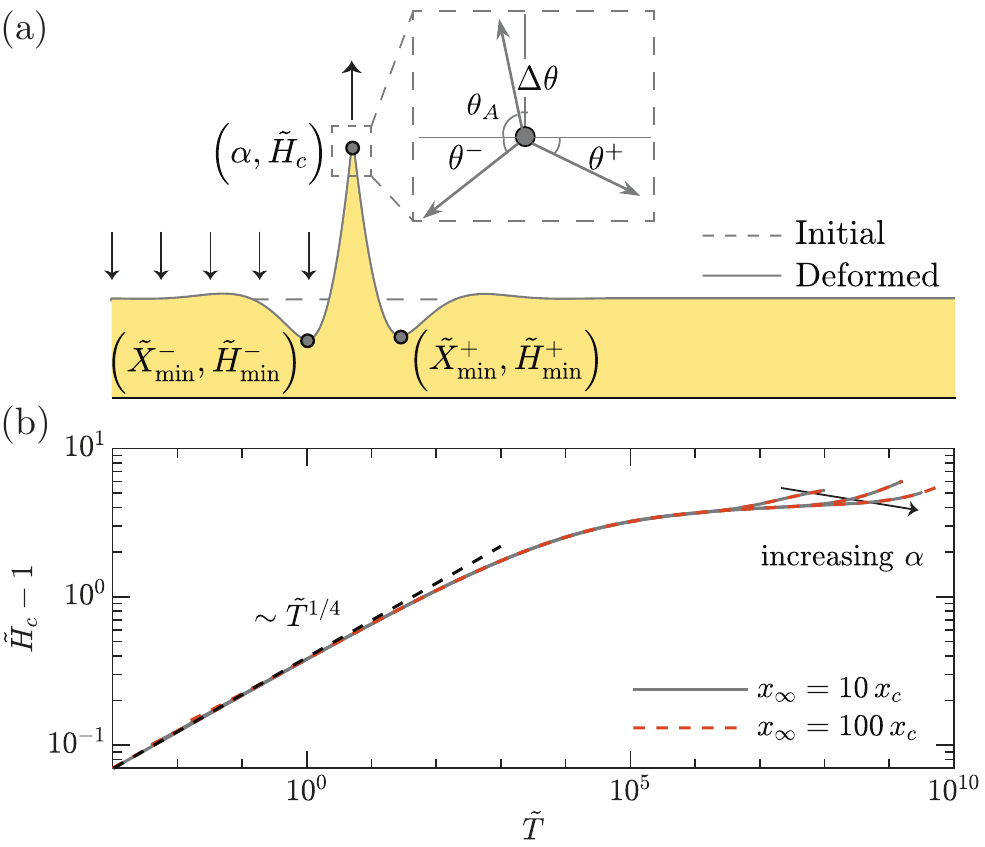}
    \caption{Numerical results for the early time evolution of the thin film profile. (a) Schematic of the deformation: two narrow neck regions emerge from the triple line and move away from it. The skirt volume is calculated dynamically as the volume contained in the wedge above the line $\eH=\eH(\As\xinf/x_c)$; at early times, $\eH(\As\xinf/x_c)\approx1$, which is the  dashed line labelled `initial'.  The zoom-in view of the skirt apex  illustrates the tilting of the Neumann triangle. (b) The early-time evolution of the change in skirt height, $\Tilde{H}_{c}-1$. Numerical results are shown by solid curves for film aspect ratios $\As=100, 200, 400$ (curves) with different values of the system size indicated by line style: $\xinf/x_c=10$ (solid grey) and  $\xinf/x_c=100$ (dashed orange). The dashed black line indicates the prediction, \eqref{eq:FT_skrthgt}, that comes from the early-time similarity solution. Here $\Bo=10^{-2}, \AvdW = \alpha^4/(10^{12})$ so that increasing $\As$ corresponds to increasing $h_0$.}    \vspace{0.5cm}
    \label{fig:DynProfile}
\end{figure}

Unless stated otherwise, the parameters used in numerical simulations are $\As=100$, $\Bo=10^{-2}$, and $\AvdW=10^{-4}$ ---  Table \ref{tab:table1} shows that these values are  typical of experiments. However, to mimic the effect of varying the film thickness, we vary $\alpha$ but choose $\AvdW\propto\alpha^4$. Numerical results for the evolution of the triple line height are shown in fig.~\ref{fig:DynProfile}b with $\alpha$ varying to mimic this film thickness variation. On the time scale of these simulations, the results show no sign of dependence on the system size, $\xinf$, and only a weak dependence on the film thickness $\alpha$, even at very late times. We therefore begin by focussing on understanding these very early stages of the motion.

\subsubsection{Linearized analysis}

In the very early stages of the motion, the system evolves away from the flat initial condition \eqref{eq:IntCd_Early}. To study this evolution, we let $\eH(\eX,\eT)= 1 + \eta(\eX,\eT)$ where $|\eta|\ll1$ and linearize \eqref{eq:EvlEq_Early} to give
\begin{equation}
    \frac{\partial \eta}{\partial \eT} = \frac{\partial^2\ePi}{\partial \eX^2} -\frac{\partial^4 \eta}{\partial \eX^4}. 
    \label{eq:LinEvol}    
\end{equation}
A scaling analysis of this equation suggests that there is an evolving length scale $\eX\sim \eT^{1/4}$ \cite{Hack2018}. In the very early stages of the motion, therefore, the effect of the second meniscus (at $X=-\As$ for the meniscus shown in fig.~\ref{fig:DynProfile}) and the far-field boundary at $\eX=\As \Xinf$ are expected to be negligible --- in agreement with the numerical results of fig.~\ref{fig:DynProfile}b. Moreover, the role of the `pushing' component of $\ePi(\eX)$ is expected to be minimal (since the  term in $\ePi$ that represents the pushing of the drop is less singular than the Dirac $\delta$-function that represents the pulling of the droplet--vapor interface). We therefore consider briefly the simplest problem of a single meniscus pulling up on the surface, corresponding to $\ePi(\eX)=-\delta(\eX-\As)$. Using the symmetric definition of the Fourier Transform
\begin{equation}
    \hat{\eta}(k,\eT)=\frac{1}{\sqrt{2\pi}}\int_{-\infty}^{\infty}\eta(\eX,\eT)e^{-ik\eX}\,\dd \eX,
\end{equation}
we find that \eqref{eq:LinEvol} is transformed to
\begin{equation}
    \frac{\partial \hat{\eta}}{\partial \eT}+k^4\hat{\eta}=\frac{1}{\sqrt{2\pi}}k^2e^{-ik\As}, 
\end{equation} 
which has solution
\begin{equation}
    \hat{\eta}(k,\eT)=\frac{e^{-ik\As}}{\sqrt{2\pi}k^2}\left(1-e^{-k^4\eT}\right).
    \label{eq:SolFT}
\end{equation}
Inverting \eqref{eq:SolFT} gives a similarity solution for the perturbation to the flat interface shape
\begin{equation}
    \eta(\eX,\eT)=\frac{1}{{2\pi}}\eT^{1/4}f(\xi),
\end{equation}
where $\xi=(\eX-\As)/\eT^{1/4}$ is the similarity variable and 
\begin{equation}
    f(\xi)=2\Gamma\left(\frac{3}{4}\right) {}_1F_3\left(-\frac{1}{4};\frac{1}{4},\frac{1}{2},\frac{3}{4};\frac{\xi^4}{256}\right)+\Gamma\left(\frac{5}{4}\right)\xi^2{}_1F_3\left(\frac{1}{4};\frac{3}{4},\frac{5}{4},\frac{3}{2};\frac{\xi^4}{256}\right)-\pi|\xi|,
    \label{eqn:SimSoln}
\end{equation} 
with ${}_1F_3(\cdot)$  the generalized hypergeometric function \cite{abramowitz64}. 

\begin{figure}[htp]
    \centering
    \includegraphics[width=9cm]{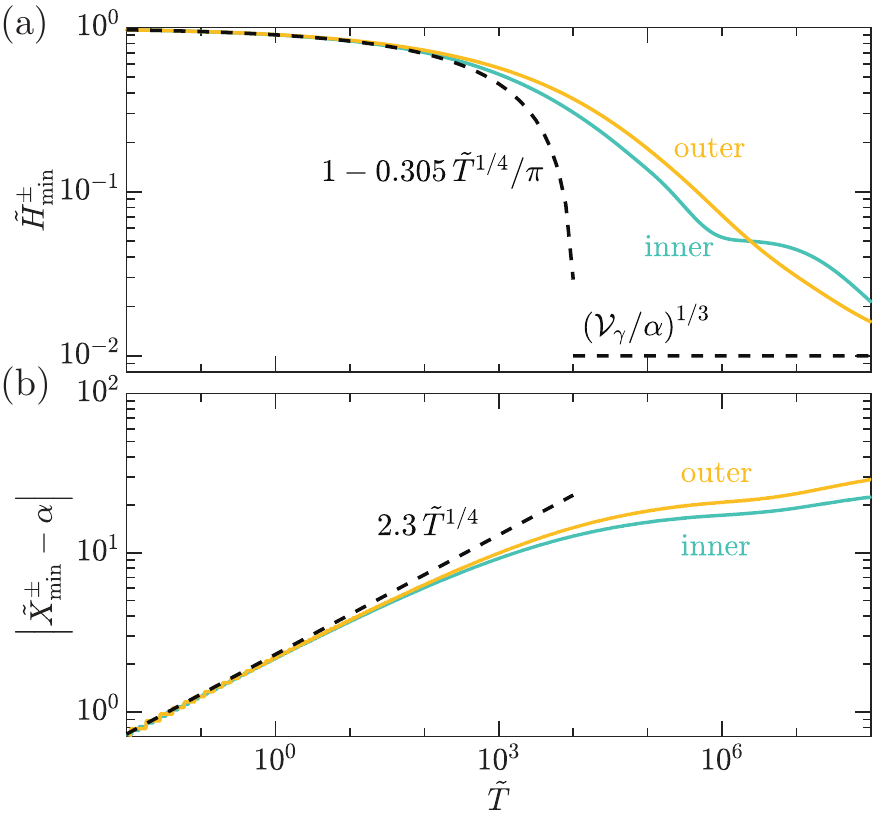}
    \caption{Early-time evolution of (a) the minimum thickness  and (b) the lateral position of the neck region that forms in the early stages of the motion. The dashed black lines are the predictions of the early time similarity solution \eqref{eq:FT_dimple}, while the solid curves show numerical results with the labels `outer' and `inner' referring to the neck regions in the regions $\eX>\As$ and $\eX<\As$, respectively. Here $\As=100$.}     \vspace{0.5cm}
    \label{fig:SkirtDimple}
\end{figure}

From the analytical expression \eqref{eqn:SimSoln} we find that $f(0)=2{\Gamma(3/4)}$, immediately giving the growth of the capillary ridge with time. The position of the necks nearest to the triple line requires the numerical determination of the position of the minima of  $f(\xi)$; these are found to occur at $\xi=\xi_*\approx\pm2.30$ with $f(\xi_*)\approx-0.610$. Altogether, this similarity solution predicts that the height of the triple line evolves according to
\begin{equation}
    \Tilde{H}_{c} \approx 1+\frac{\Gamma(3/4)}{\pi} \eT^{1/4}.
    \label{eq:FT_skrthgt}
\end{equation}
while the minima have location and film thickness
\begin{equation}
    {\Tilde{X}_\mathrm{min}^\pm} \approx\As \pm 2.30\,\eT^{1/4},\quad {\Tilde{H}_\mathrm{min}^\pm} -1\approx - \frac{0.305}{\pi}\,\eT^{1/4}.
    \label{eq:FT_dimple}
\end{equation} Consideration of the function $f(\xi)$ in \eqref{eqn:SimSoln} shows that there are two local maxima in the film height, $\eH_\mathrm{max}^\pm$, whose location and film thickness are
\begin{equation}
    {\Tilde{X}_\mathrm{max}^\pm} \approx\As \pm 5.81\,\eT^{1/4},\quad {\Tilde{H}_\mathrm{max}^\pm} -1\approx  0.009\,\eT^{1/4},
    \label{eq:FT_max}
\end{equation} while stagnation points between these minima and maxima lie at
\begin{equation}
    \tilde{X}_{\mathrm{s.p.}}^{\pm}\approx \As\pm4.592\tilde{T}^{1/4}.
    \label{eqn:StagPtsEarlyT}
\end{equation} While many local maxima, minima and stagnation points exist in this similarity solution (and indeed throughout the evolution), we shall see that these first ones beyond the triple line are particularly important in the ensuing dynamics. We will refer to these first local maxima as the `bumps' in what follows, while the minima are referred to as `necks'.

The above calculation gives some useful predictions for the early time evolution that can be tested by comparison with the numerical results. Figure \ref{fig:DynProfile} shows that the height of the triple line does indeed follow the prediction of \eqref{eq:FT_skrthgt}. Similarly, the predictions of \eqref{eq:FT_dimple} are borne out by the numerical results shown in fig.~\ref{fig:SkirtDimple}.

\begin{figure}[htp]
    \centering
    \includegraphics[width=9.5cm]{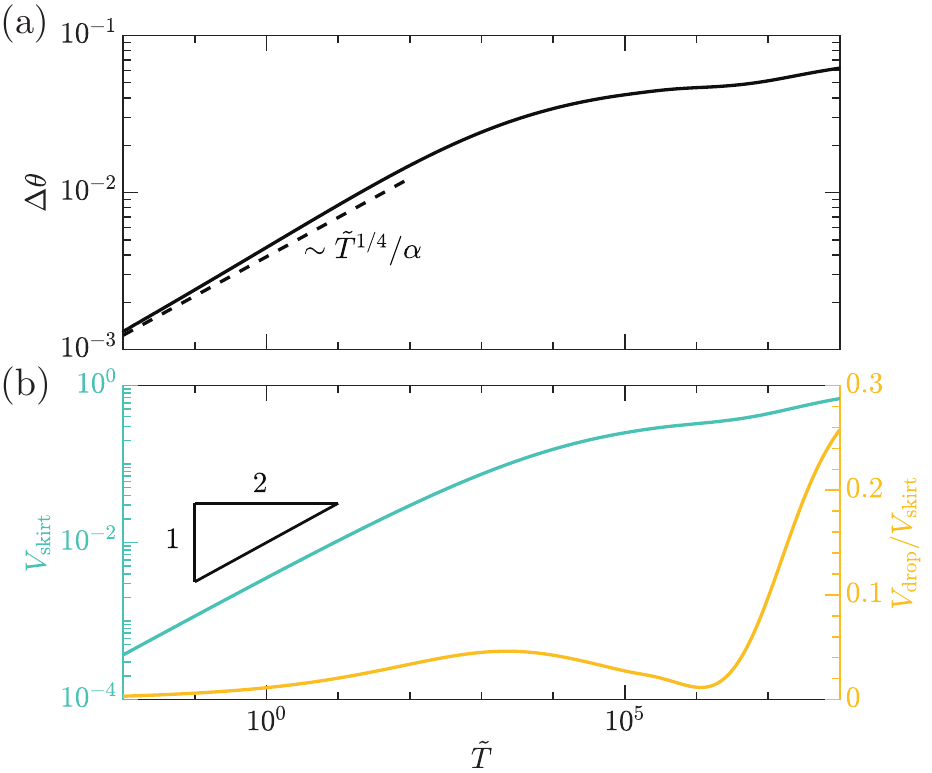}
    \caption{Early-time evolution of (a) the induced rotation angle, $\dt$,  and (b) the  volume, $\Vskirt$,  of the skirt. In (a) the asymptotic prediction  \eqref{eq:HorF} is shown by the dashed line. In (b), $\Vskirt$ is the volume of liquid near the triple line, and is defined by $\Vskirt(t)=\int_{\xmin^-}^{\xmin^+}[h(x,t)-h(\xinf,t)]~\dd x/(x_c h_0)$. $\Vdrop(t)$ is the rescaled volume of the liquid squeezed out beneath the drop, i.e., $\Vdrop(t)=\int_{0}^{x_c}[h_0-h(x,t)]~\dd x/(x_c h_0)$.}    \vspace{0.5cm}
    \label{fig:RotVol}
\end{figure}

While this similarity solution gives some insight into the early stages of the motion, its neglect of both the second meniscus (at $\eX=-\As$) and of the pushing pressure mean that its predictions remain purely symmetric throughout, clearly limiting its utility in describing later stages of the motion. To answer any questions about how the asymmetry at the triple line evolves dynamically requires these ingredients to be put back into the model, i.e.~for the full form of $\ePi(\eX)$ to be used. The same analysis can be followed with the complete loading pressure, i.e.~$\ePi(\eX)=\ind(\eX)/\As-[\delta(\eX+\As)+\delta(\eX-\As)]$,  by taking Fourier transforms of \eqref{eq:LinEvol}.

In this case, the solution is
\begin{equation}
    \hat{\eta}(k,\eT)=\sqrt{\frac{2}{\pi}}\left(\frac{\cos{\As k}}{k^2}-\frac{\sin{\As k}}{\As k^3}\right)\left(1-e^{-k^4\eT}\right).
    \label{eq:SolFT2}
\end{equation} We are not able to give a closed form for the inverse of \eqref{eq:SolFT2}, and note that the introduction of a length scale (the separation between the two menisci), breaks the similarity form of the solution. However, it is possible to make progress in understanding the evolution of the quantity of most interest, which  is the slope of the profile surrounding the triple line. In particular, we find that the slope of the interface deformation throughout is given by
\begin{equation}
    \frac{\partial \eH}{\partial \eX}=    \frac{i}{\pi}\int_{-\infty}^{\infty}\left(\frac{\cos{\As k}}{k}-\frac{\sin{\As k}}{\As k^2}\right)\left(1-e^{-k^4\eT}\right)e^{ik\eX}\,\dd k.
    \label{eq:SlpFT}
\end{equation}

This expression can be used to calculate the asymmetry in the profile either side of the triple line and induces a  tilt $\dt=\pi/2-\theta_A$ in the  vertical force applied to the film to satisfy horizontal force balance. Though our numerics neglect this tilt, its required magnitude can be calculated assuming $\dt\ll1$; using the linearized version of \eqref{eq:RotAng} with \eqref{eq:SlpFT} we find that
\begin{equation}
    \dt \approx \frac{1}{2}\left(\left.\frac{\partial \eH}{\partial \eX}\right\vert_{(\As^-,\,\eT)}+\left.\frac{\partial \eH}{\partial \eX}\right\vert_{(\As^+,\,\eT)}\right) = \frac{\Gamma(3/4)}{\pi\As}\eT^{1/4},
    \label{eq:HorF}
\end{equation}
at early times. This result is in good agreement with our numerical simulations, as shown in fig.~\ref{fig:RotVol}a.


\section{Dynamics: Beyond early times} 

The preceding analysis was predicated on the assumption that the relevant time scale was that on which the initial film thickness varies, since we envisage this is the limiting factor in the first moments after a droplet is deposited. However, we also know that the ultimate equilibrium is well-described by using a non-dimensionalization based on the drop size, $x_c$, in the horizontal direction and the initial film thickness $h_0$ in the vertical direction. A scaling analysis of  Reynolds' equation then gives us that the relevant time scale is $\tau\propto \mu x_c^4/(\gamma h_0^3)$. Typically this time scale is on the order of  $10^{2}\mathrm{~s}$,  which is orders of magnitude longer than the early time scale $\mu h_0/\gamma$, but comparable to that over which experiments might be expected to occur. 

\begin{figure}[!htp]
    \centering
    \includegraphics[width=16cm]{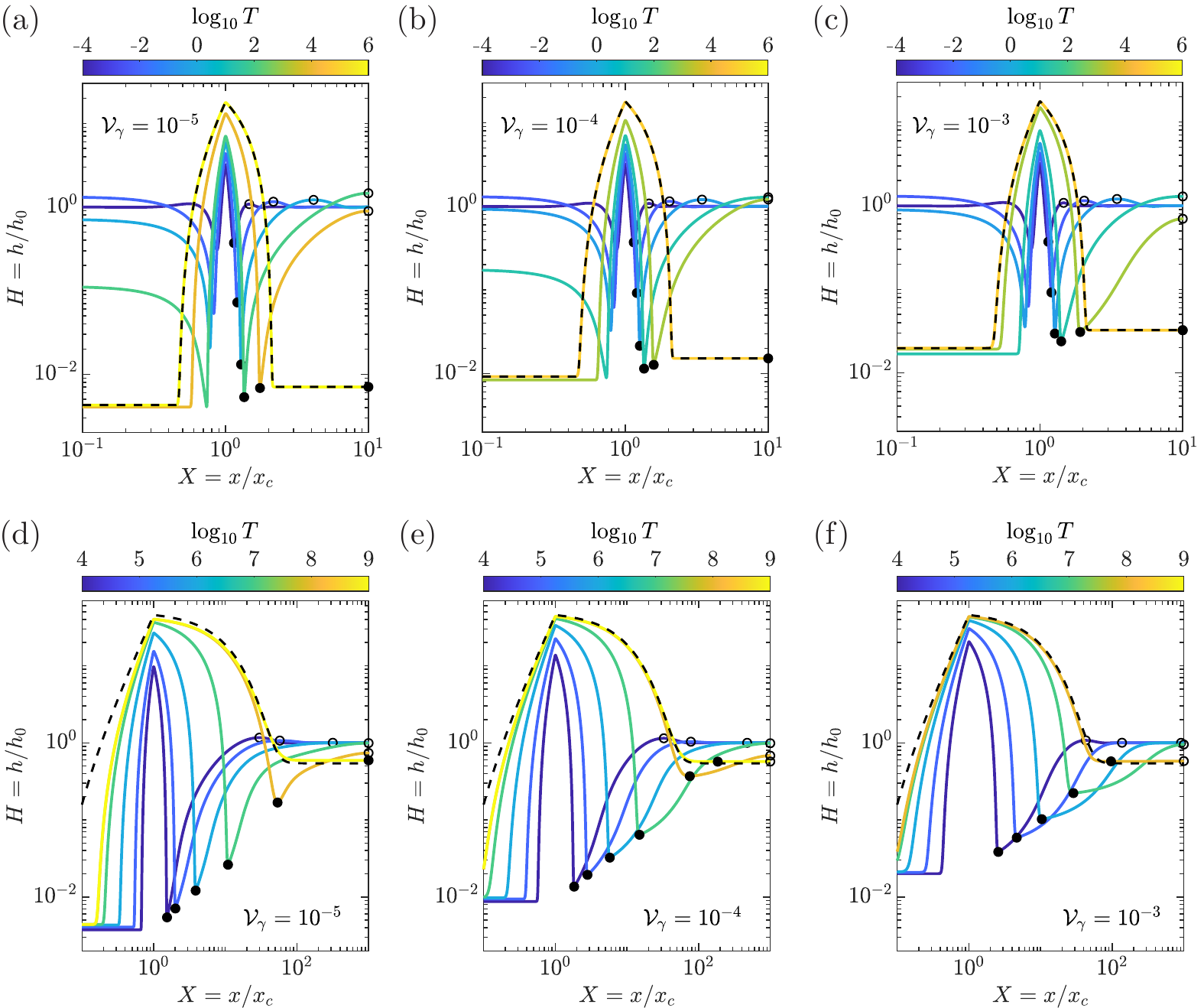}
    \caption{The evolution of film thickness for an initially flat oil film subjected to the spatially non-uniform pressure profile of the droplet. Here the effect of the van der Waals parameter $\AvdW$ on the time scale of evolution may be seen as $\AvdW$ varies from $10^{-5}$ (in a,\,d) to $10^{-4}$ (in b,\,e) and $10^{-3}$ (in c,\,f). Here $\Bo^{-1}=\As=10^2$ in all calculations; in (a)-(c) $\Xinf=10$ while in (d)-(f)  $\Xinf=10^3$. Note that the film profile ultimately reaches very close to the equilibrium solution (dashed black curves), but that the time taken to do so is extraordinarily long even for $\Xinf=10$ ($T\gtrsim10^4\sim 10\mathrm{~days}$ with a typical $\AvdW=10^{-4}$). Throughout, line colour is used to encode dimensionless time as in the associated colorbar. Black filled and open markers are used to indicate the local minima (necks) and maxima (bumps), respectively.}    \vspace{0.5cm}
    \label{fig:EvolvingProfiles}
\end{figure}

\subsection{Non-dimensionalization}

We rescale lengths in the same manner as the static problem, i.e.~we let $X=x/{x_c}$ and $H=h/{h_0}$; in this way, the natural timescale discussed above arises, though we formally let
\begin{equation}
    \tau = \frac{3\mu x_c^4}{\gamma h_0^3}=\alpha^4\tau_\ast
\end{equation} for notational convenience. We find that the dimensionless form of the problem reads
\begin{equation}
    \frac{\partial H}{\partial T} = \frac{\partial}{\partial X}\left[H^3\frac{\partial\Pi}{\partial X} -H^3\frac{\partial^3 H}{\partial X^3}+\left(\frac{3\AvdW}{H}+ \Bo\,H^3\right)\frac{\partial H}{\partial X}\right],
    \label{eq:EvlEq_Late}    
\end{equation}
subject to
\begin{gather}
    H(X,0)=1,
    \label{eq:IntCd_Late}\\
    H_X(0,T)=H_X(\Xinf,T)=H_{XXX}(0,T)=H_{XXX}(\Xinf,T)=0,
    \label{eq:BC_Late}    
\end{gather}
for $0\le X \le \Xinf$, where $\Pi=\As\ind(X)-\As\delta(X-1)$.

\subsection{Numerical results}

Although the new  time scale introduced to study the late time dynamics is significantly longer than that relevant at early times, allowing the system to reach very close to equilibrium still requires long dimensionless times. We use the same numerical technique as already described. However, a key difficulty in achieving this numerically is the very small horizontal scales over which the forcing pressure $\Pi(X)$ changes, compared to the very long length scales on which the interface shape is changing. To address this, we solve the problem \eqref{eq:EvlEq_Late}--\eqref{eq:BC_Late} numerically with a numerical scheme that smooths the indicator and Dirac-$\delta$ functions present in the forcing pressure over a small length scale; this scale is chosen to be small enough to resolve the early time dynamics already discussed, but large enough that we can reach close to the ultimate equilibrium. (See Appendix \ref{sec:Numerics} for details.)

\begin{figure}[!htp]
    \centering
    \includegraphics[width=14cm]{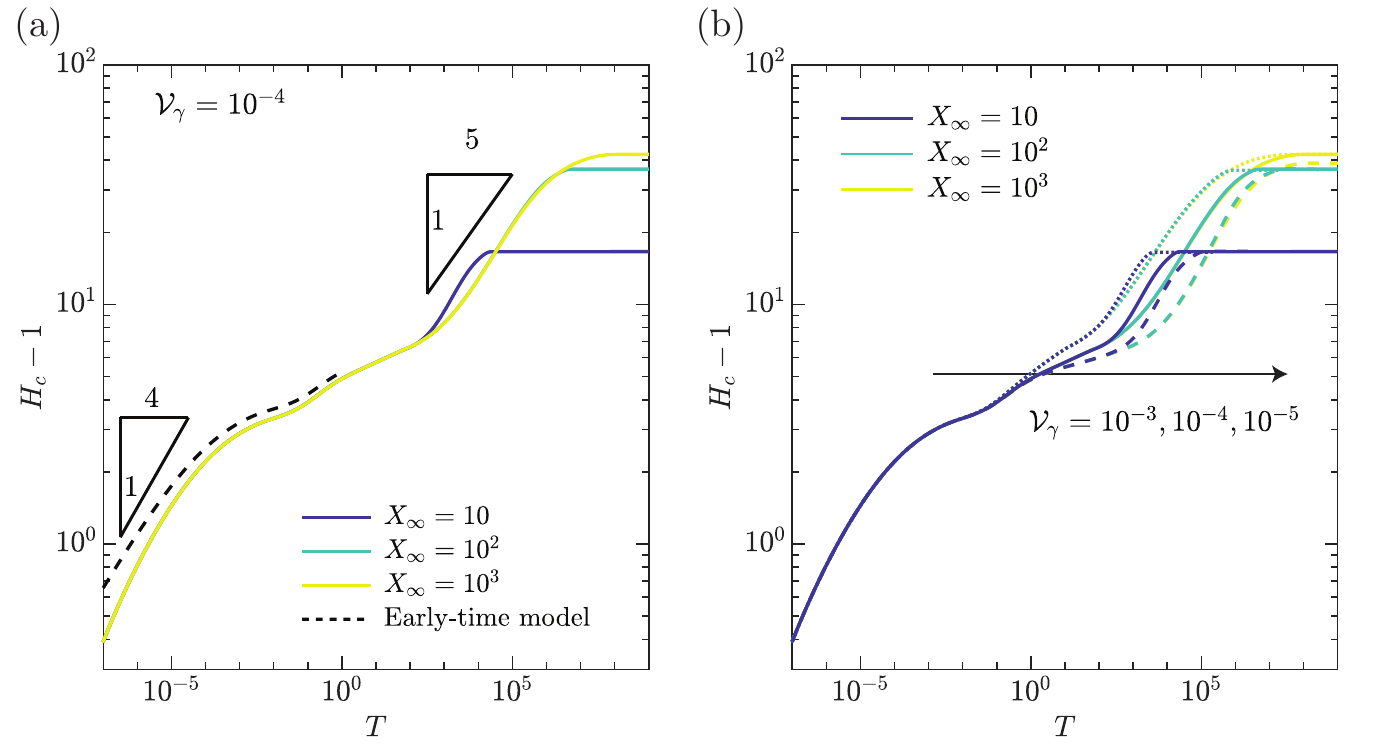}
    \caption{(a) Late-time evolution of the skirt height with a typical value, $\AvdW=10^{-4}$, and different system sizes (denoted by line colour, as described in the legend). Note that dashed black curves are the results of the early-time analysis (where the true $\delta$-function forcing is implemented) while solid curves show results with the smoothed forcing (see main text). The dynamic system is not sensitive to the system size $\Xinf$ until $T\sim10^{3}$, at which point the film height at the outer boundary begins to change for the smallest system, $\Xinf=10$. (b) Late-time evolution of the skirt height with $\AvdW$ varying: $\AvdW=10^{-3}$ (dotted), $\AvdW=10^{-4}$ (solid) and $\AvdW=10^{-4}$ (dashed) for various plate sizes (as indicated by line colour). In all calculations $\As=\Bo^{-1}=10^2$.}    \vspace{0.5cm}
    \label{fig:HeightnRot_Late}
\end{figure}

Figure \ref{fig:EvolvingProfiles} shows the evolution of the film thickness profile towards equilibrium interface shape found in \S\ref{sec:NumsMainText}. This validates the smoothing of the pressure profile $\Pi(X)$ that is used for numerical convenience, but also illustrates another important point: the skirt region grows extremely slowly. Further validation of the numerical smoothing of the triple line discontinuity is provided by the evolution of the meniscus height and wedge rotation as functions of time, see fig.~\ref{fig:HeightnRot_Late}a.

The numerical results in fig.~\ref{fig:EvolvingProfiles} show  features that are distinct from the behaviour already discussed at very early times. Of particular significance is the behaviour of the local maximum,  $(\Xmax,\Hmax)$, beyond the neck region centred on  $(\Xmin,\Hmin)$ that we have already discussed: this maximum  plateaus at a constant height even before the size of the system, $\Xinf$, becomes important. This change in behaviour (compared to the continued growth of the maximum observed in the early-time similarity solution) is the signature of a transition to one of several regimes that the system moves through. We now focus on these regimes, but a schematic summary is given in fig.~\ref{fig:RegimesDiagram}. 

\begin{figure}[!htp]
    \centering
    \includegraphics[width=9cm]{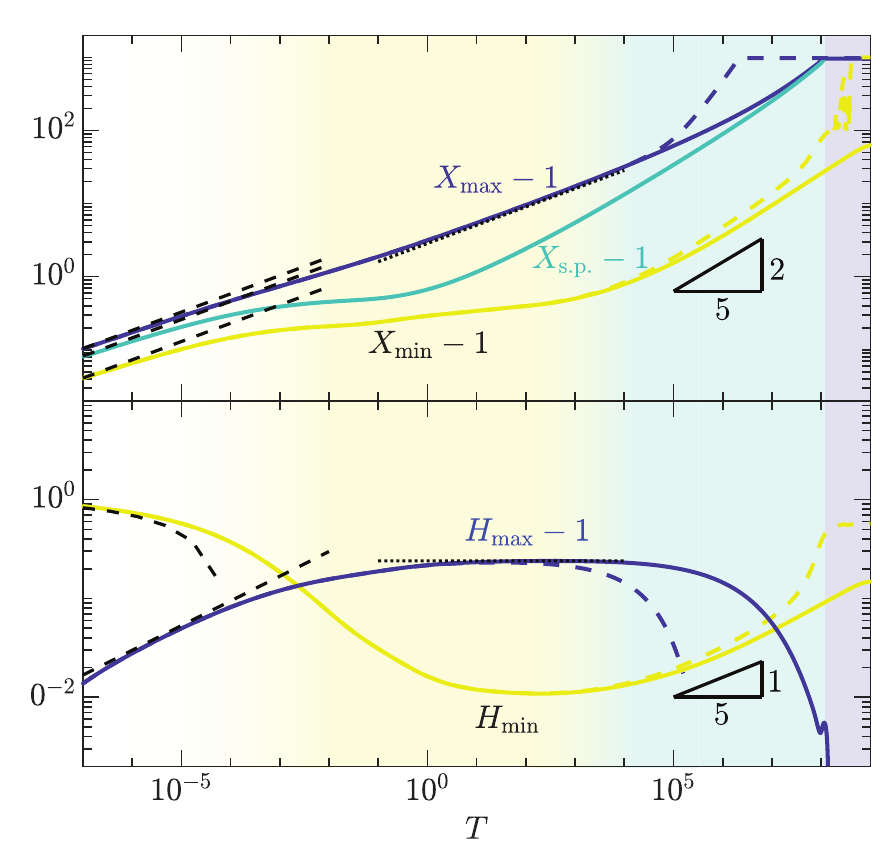}
    \caption{Time evolution of the properties of the local minimum (the neck region, yellow curves) and the maximum (the bump region, blue curves). Each of these regions evolves through several different behaviours. A key feature is the position of the stagnation point located between them --- $X_{\mathrm{s.p.}}$ is the point at which the fluid flux vanishes, $\partial P/\partial X=0$, and is shown by the cyan curve. Here, $\As=100$ and $\AvdW=\As^4/10^{12}=10^{-4}$; solid curves use $\Bo=0$ while  dashed curves used $\Bo=10^{-2}$. Different background shading is used to represent early times, early--intermediate times, late--intermediate times, and late times (from left to right). Black dashed and dotted curves are plotted using the corresponding analytical expressions, which are summarized in the first two panels of fig.~\ref{fig:RegimesDiagram}. }     \vspace{0.5cm}
    \label{fig:DimpleBumpEvolution}
\end{figure}

To understand these regimes, we first focus on numerical results for $\As =10^2,\AvdW=10^{-4}$ with zero gravity, $\Bo=0$,  and the largest system size, $X_\infty=10^3$. The large system size and neglect of gravity simplify the problem slightly and allow us to extend the time scale over which various phenomena are observed for as long as possible. (We find that smaller system sizes show the same behaviour but only over shorter time periods, while the effect of gravity is negligible until the very latest stages of the motion, as can be seen by comparing the solid curves in fig.~\ref{fig:DimpleBumpEvolution}, for which $\Bo=0$, with the dashed curves, for which $\Bo=10^{-2}$. We revisit the effects of finite Bond number in \S\ref{sec:Gravity}, but note that even with $\Bo=0$ an equilibrium is ultimately reached since van der Waals forces also play the role of a restoring force. We shall therefore simplify the analysis until that point by assuming $\Bo=0$ in what follows.) 

An important feature of the results shown in figure \ref{fig:DimpleBumpEvolution} is the evolution of the position of the stagnation point between the external film minimum and the external maximum.  The early time similarity solution shows that this stagnation point is initially located close to the `bump' region, as shown by comparing eqns \eqref{eq:FT_max} and \eqref{eqn:StagPtsEarlyT}. However, fig.~\ref{fig:DimpleBumpEvolution} shows that it moves towards the `neck' region,  before ultimately moving again towards the bump. We shall see that this movement of the stagnation point delineates different phases of the motion.


\subsection{Early--intermediate times}

\begin{figure}[!htp]
    \centering
    \includegraphics[width=14.5cm]{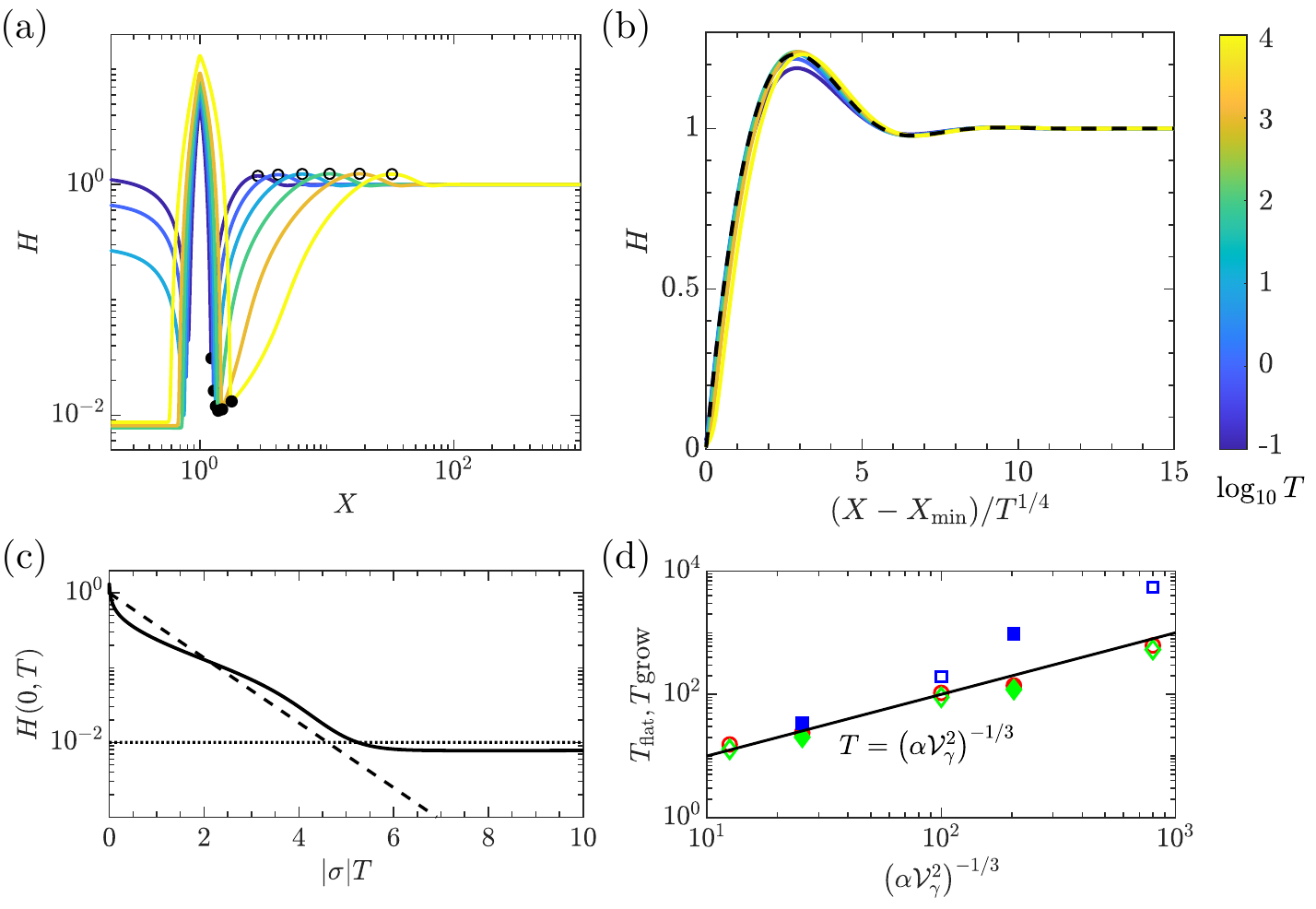}
    \caption{Evolution of the film profile at early--intermediate times in which a `bump' of fixed height $\Hplateau$ propagates away from the droplet.  (a) The unscaled profile showing the neck at $\Xmin$ (filled black  circle) and the propagating bump (open black circle). (b) The raw data of (a) rescaled according to \eqref{eqn:InterTimeSimSoln} agrees well with the numerical solution of the intermediate-time similarity equation \eqref{eqn:BumpPropSimSoln} (dashed curve). (In (a) and (b) curves are colored according to time as in the colorbar to the right.) (c) The numerically determined film height at the centre of the drop, $X=0$, as a function of $|\sigma|T$ (solid curve) compared to the the prediction of the linear stability analysis, $H(0,T)\propto\exp{(-\sigma T)}$, with $\sigma$ given by \eqref{eqn:EarlyInterTime} calculated with $\AvdW=10^{-4}$, $\As=10^2$ (dashed line). At late times, the numerics are close to the equilibrium thickness $\Heqm=(\AvdW/\As)^{1/3}$ (dotted line).   (d) The time needed for the film beneath the droplet to become flat ($T_\mathrm{flat}$, red circles) and for the film thickness beneath the two neck regions to stop decreasing and begin growing (denoted $T_\mathrm{grow}$ and shown by green diamonds for the inner neck and blue squares for the outer neck). Results are shown for $\AvdW= \As^4/10^{12}$ (hollow symbols) and stronger van der Waals forces, $\AvdW=\As^4/(2\times10^7)$, (filled symbols) where $\As$ varies from 50 to 200. The solid line shows the scaling prediction $T\propto\sigma^{-1}$ from the stability analysis \eqref{eqn:EarlyInterTime} with prefactor such that $T=(\As\AvdW^2)^{-1/3}$ or, in dimensional form, $t=3\mu x_c^{7/3}\gamma^{-1/3}A^{-2/3}$, independent of the initial film thickness.}
    \vspace{0.5cm}
    \label{fig:EarlyIntermediateTimes}
\end{figure}

The early time predictions become invalid when the vertical deformation  reaches the same order of magnitude as the initial film thickness, i.e.~when $\As T^{1/4}\sim1$ or $T\sim\As^{-4}$. Beyond this point, the evolution of the film  thickness beneath the neck and its position slow down significantly but, perhaps surprisingly, the local maximum (the `bump') still evolves horizontally following the early-time scaling $\Xmax\sim T^{1/4}$. However, two important differences from the early-time behaviour can be seen in fig.~\ref{fig:DimpleBumpEvolution}: firstly the prefactor $\Xmax/T^{1/4}$ decreases from that given in \eqref{eq:FT_max} and, secondly, the height of the bump, $\Hmax$ does not continue increasing, but rather plateaus at a new value, $\Hplateau$; an estimate of the time $\Tplateau$ needed for this to occur may be obtained from \eqref{eq:FT_max} and the observation that $\Hplateau\sim 1$ giving $\Tplateau\sim \As^{-4}$. For $T\gtrsim \Tplateau$, the bump appears to propagate outwards with an approximately constant amplitude $(\Hplateau-1)$ that is maintained for a long period (fig.~\ref{fig:EarlyIntermediateTimes}a). This change in behaviour is associated with the relative proximity of the neck region and stagnation point: because the neck region has reached a very small thickness, the flux of fluid through it into the skirt region is very small. The neck region therefore separates the droplet and skirt regions from what happens beyond the neck region, in which the liquid film has to interpolate between a small thickness region to its left (the neck) and a relatively large thickness region to its right (the film). This is then reminiscent of the capillary healing/levelling problem  for a thin film that has been considered recently \cite{Benzaquen2014,Zheng2018}; we are able to make use of these results to understand this behavior, as we now show.

\subsubsection{Intermediate time similarity solution for the bump}

To understand the propagating, but constant height, bump regime and make the analogy with capillary healing more concrete, we note that van der Waals forces may be neglected in this region of the film (because the bump itself is well beyond the scale at which van der Waals forces play a role). The evolution of the film thickness can then be approximately described by
\begin{equation}
    \frac{\partial H}{\partial T} = -\frac{\partial}{\partial X}\left(H^3\frac{\partial^3 H}{\partial X^3}\right).
    \label{eqn:BumpPropagate}
\end{equation}

We seek a similarity solution of \eqref{eqn:BumpPropagate} of the form \begin{equation}
    H(X,T)=\eta(\xi),\quad \mathrm{with}\quad \xi=(X-\Xmin)/T^{1/4}.
    \label{eqn:InterTimeSimSoln}
\end{equation} The rescaled equation reads
\begin{equation}
    \frac{\partial}{\partial\xi}\left(\eta^3\frac{\partial^3\eta}{\partial\xi^3}\right)-\left(\frac{\xi}{4}+{\dot{X}_\mathrm{min}T^{3/4}}\right)\frac{\partial\eta}{\partial\xi}=0,
    \label{eqn:BumpPropSimSolnStep1}
\end{equation}
subject to boundary conditions
\begin{equation}
    \eta\to0,~\frac{\partial\eta}{\partial\xi}\to0 \text{~as~}\xi\to0\quad\text{and}\quad
    \eta\to1,~\frac{\partial\eta}{\partial\xi}\to0 \text{~as~}\xi\to+\infty.
    \label{eq:LateTimeSimilarity_BC}
\end{equation} (In this phase of the motion, the film thickness beneath the neck, $\Hmin$, is controlled by van der Waals forces so that  $\Hmin\sim (\AvdW/\As)^{1/3}\ll1$; we take $\Hmin=\eta(0)\approx0$ for simplicity.)

Equation \eqref{eqn:BumpPropSimSolnStep1} is not of similarity form because of the time dependence of the neck position $\Xmin$; however, since in this stage of the motion the neck location evolves  more slowly than at early times (where  $\Xmin-1\sim T^{1/4}$), we take $\dot{X}_\mathrm{min}T^{3/4}\ll1$, and \eqref{eqn:BumpPropSimSolnStep1} simplifies to become
\begin{equation}
    \frac{\dd}{\dd\xi}\left(\eta^3\frac{\dd^3\eta}{\dd\xi^3}\right)-\frac{\xi}{4}\frac{\dd\eta}{\dd\xi}=0.
    \label{eqn:BumpPropSimSoln}
\end{equation} Equation \eqref{eqn:BumpPropSimSoln}  is precisely the equation describing the early-time similarity solution of an initial step of liquid relaxing under surface tension (see eqn A2 of ref.~\cite{Zheng2018}, for example).

The numerical solution of \eqref{eqn:BumpPropSimSoln} subject to \eqref{eq:LateTimeSimilarity_BC} is plotted as the dashed curve in fig.~\ref{fig:EarlyIntermediateTimes}. (We use a finite domain $[0,\xi_\infty]$, but find that $\xi_\infty=15$ is sufficient to obtain good convergence of the numerical solution.)  From this numerical solution, we can readily calculate the position and height of the bump from the rescaled profile $\eta(\xi)$; we find that the maximum is $\Hplateau\approx1.25$, and occurs at $\xi\approx2.84$. This calculation provides direct predictions for the evolution of the bump in the early--intermediate times stage, which are summarized in fig.~\ref{fig:RegimesDiagram}b and compare very well to the numerical results of the dynamic problem (dotted lines in the yellow domain of fig.~\ref{fig:DimpleBumpEvolution}). (The solution shown in fig.~\ref{fig:EarlyIntermediateTimes} is similar to that calculated by \citet{Zheng2018}, see their fig.~16. However, in our solution there is no pre-wetted layer as $\xi\to-\infty$ and so the height of the maximum is slightly larger.)

\subsubsection{Beneath the drop as the bump propagates}

The propagation of the bump was predicated on the assumption that the liquid film is effectively split in two by the stagnation point close to the neck. This then begs the question: what is happening beneath the droplet during this time? At this point in the evolution, the region beneath the droplet is a concave dimple region: the pressure within this dimple is positive and so this dimple drains into the skirt region (through the internal neck region, see fig.~\ref{fig:RegimesDiagram}b). This drainage in turn flattens the dimple out, ultimately reaching the uniform equilibrium thickness at which the van der Waals pressure balances the Laplace pressure of the droplet, namely $\Heqm=(\AvdW/\As)^{1/3}$. (Note, however, that (i) the equilibrium shape beneath the droplet may not ultimately have such a large flat spot, because of the finite size of the droplet and (ii) the ultimate thickness of the layer beneath the drop may be slightly less than $\Heqm$ because of the additional negative pressure in the skirt.)

To understand this flattening motion and the timescale on which it occurs, we consider the linear stability of the homogeneous thickness (i.e.~flat) region by substituting $H(X,T)=\Heqm+\delta f(X)\exp(\sigma T)$ (with $\delta\ll1$ arbitrary) into \eqref{eq:EvlEq_Late} and linearizing. This procedure gives
\begin{equation}
    \sigma f\approx-\Heqm^3 f''''+3\AvdW f''/\Heqm.
\end{equation} which is to be solved subject to symmetry boundary conditions, $f'(0)=f'''(0)=0$, and zero displacement and pressure boundary conditions, $f(1)=f''(1)=0$;  the solutions are of the form $f=\cos\bigl[(n+\tfrac{1}{2})\pi X\bigr]$ and so the slowest decaying mode has $n=0$ and decay rate
\begin{equation}
    |\sigma|=-\sigma=\frac{\pi^4\Heqm^3}{16}+\frac{3\pi^2\AvdW}{4\Heqm}\approx\frac{3\pi^2}{4}\As^{1/3}\AvdW^{2/3}
    \label{eqn:EarlyInterTime}
\end{equation} since $\Heqm=(\AvdW/\As)^{1/3}$, $\AvdW\ll1$, and $\As\gg1$. We therefore expect the time scale for the collapse of the dimple towards the flat spot beneath the droplet should scale as $|\sigma|^{-1}\propto \left(\As\AvdW^2\right)^{-1/3}$. This scaling is confirmed by plotting the film height at $X=0$ as function of time in numerics (see fig.~\ref{fig:EarlyIntermediateTimes}c) as well as the numerical measurements of the time at which the inner dimple disappears to be replaced by the flat region (see fig.~\ref{fig:EarlyIntermediateTimes}d). (Note that the presence of the early time regime therefore requires $\As^{-4}\ll(\As\AvdW)^{-1/3}$, but this is guaranteed by the scale separation $\AvdW\ll1\ll\As$.)

\subsubsection{Deviation from self-similarity in the bump region.} Once the dimple beneath the droplet has drained to the homogeneous film thickness $\approx\Heqm$, there is no longer any further fluid available to fill the skirt region. Nevertheless, the triple line has not yet reached its equilibrium height and so the skirt must seek liquid elsewhere. The only available reservoir of fluid is within the bump region and so the system must increase the flux that flows beneath the outer neck region, which in turn requires the height there to be increased: $\Hmin$ increases, as observed numerically. This increase of $\Hmin$ in turn breaks the conditions for the constant height propagation of the bump region already discussed, and so this similarity solution must break down. In fig.~\ref{fig:EarlyIntermediateTimes}, we see that numerical simulations do indeed start to deviate from the similarity solution given by the solution of \eqref{eqn:BumpPropSimSoln} at sufficiently large $T$. This deviation can also be observed in fig.~\ref{fig:DimpleBumpEvolution} where the  predictions of the similarity solution (black dotted lines) fail to describe the  behavior of the bump at later times. With $\Hmin$ increasing, the `valve' that the outer neck region provided is released; fluid begins to flow into the skirt from beyond the droplet region again, and the stagnation point moves again towards the bump region. These changes all have consequences for the evolution of the film profile with increasing time, and so correspond to a new regime: late--intermediate times.

\subsection{Late--intermediate times}

The end of early--intermediate times is characterized by the complete drainage of the dimple beneath the droplet, which forces the release of the `valve' in the neck region: the stagnation point moves away from the neck region towards the bump and the height at the neck region increases.

\begin{figure}[!htp]
    \centering
    \includegraphics[width=14.5cm]{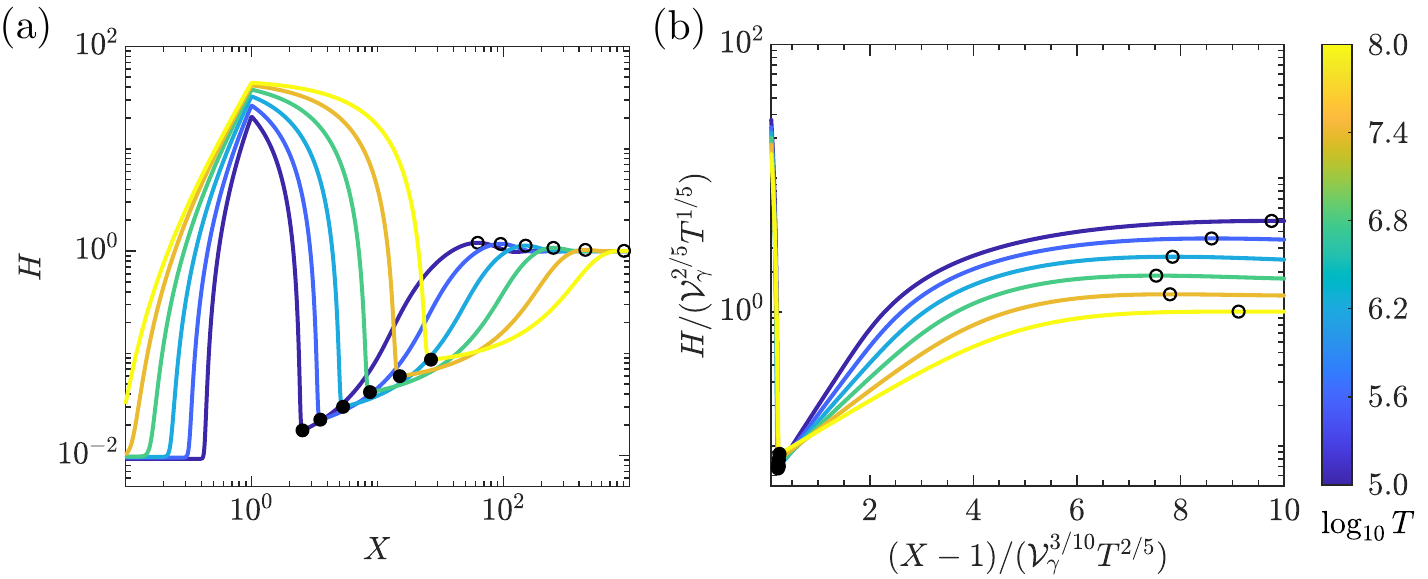}
    \caption{Evolution of the film profile at late--intermediate times. (a) Raw dimensionless and (b) scaled profiles showing the location of the neck (filled black circles) and bump (open circles) in each case. The data in (b) are rescaled according to \eqref{eqn:LateTimeRescaling} and plotted on a semi-logarithmic scale to highlight the exponential growth away from the neck region.}    \vspace{0.5cm}
    \label{fig:LateIntermediateTimes}
\end{figure}

The salient feature of this late--intermediate regime is that both the neck and the bump regions are smoothing out (fig.~\ref{fig:LateIntermediateTimes}a). We now focus on the evolution of the neck region that connects the capillary-controlled region (the skirt) to the outer, reservoir region that is mostly controlled by van der Waals forces (in the absence of gravity). We proceed by first conducting a scaling analysis of \eqref{eq:EvlEq_Late} in the outside region (so that $\dd \Pi/\dd X=0$).  We anticipate that in this phase of the motion surface tension and van der Waals forces balance, but also that the system is out of equilibrium; at a scaling level  eqn \eqref{eq:EvlEq_Late} immediately gives that the typical vertical and horizontal scales are
\begin{equation}
H\sim \AvdW^{2/5} T^{1/5},\quad X\sim \AvdW^{3/10} T^{2/5}.
\end{equation} (These scalings are analogous to those derived for thin film rupture behaviour under attractive van der Waals interactions \cite{Zhang1999}.) It is therefore natural to introduce the similarity variables
\begin{equation}
\bar{\eta}=H(X,T)/\left(\AvdW^{2/5}T^{1/5}\right),\quad \bar{\xi}=(X-1)/\left(\AvdW^{3/10}T^{2/5}\right).
\label{eqn:LateTimeRescaling}
\end{equation} With this transformation, \eqref{eq:EvlEq_Late} gives:
\begin{equation}
    \frac{1}{5}\bar{\eta}-\frac{2}{5}\bar{\xi}\frac{\dd\bar{\eta}}{\dd\bar{\xi}}=\frac{\dd}{\dd \bar{\xi}}\left(-\bar{\eta}^3\frac{\dd^3 \bar{\eta}}{\dd \bar{\xi}^3}+\frac{3}{\bar{\eta}}\frac{\dd\bar{\eta}}{\dd\bar{\xi}}\right),
    \label{eqn:EvolveLateVdW}
\end{equation} which is, as expected of a similarity solution, independent of time. Unlike previous work on the role of van der Waals forces in thin film rupture \cite{Zhang1999}, this similarity equation appears not to have appropriate exact (or even asymptotic) solutions: in particular, the far-field condition that $H\sim 1$ as $\bar{\xi}\to\infty$ is inconsistent with the similarity transformation. As such, we do not expect the numerical results of the full problem to collapse perfectly when plotted in terms of these similarity variables --- as indeed is the case (see fig.~\ref{fig:LateIntermediateTimes}b). Nevertheless, the rescaling  \eqref{eqn:LateTimeRescaling} does give a reasonable collapse of the numerical data: while some systematic evolution of the meniscus shape remains, note that the profiles shown in fig.~\ref{fig:LateIntermediateTimes}b span 3 orders of magnitude in time.) The scalings in \eqref{eqn:LateTimeRescaling} also suggest that $\Hmin\sim \AvdW^{2/5}T^{1/5}$ and $\Xmin\sim\AvdW^{3/10}T^{2/5}$, in good agreement with the numerics shown in the cyan domain of fig.~\ref{fig:DimpleBumpEvolution}. 

Although we believe that no detailed solution of \eqref{eqn:EvolveLateVdW} can be found satisfying appropriate boundary conditions for the problem at hand, we note that during this phase of the motion the region between the triple line and the neck, i.e.~$1\leq X\lesssim \Xmin$, is approximately parabolic, corresponding to constant pressure in this region. More surprising perhaps is that, immediately beyond the neck region, the film thickness grows approximately exponentially (see fig.~\ref{fig:LateIntermediateTimes}b). This can be rationalized as a constant flux (in the similarity variables) that is dominated by the van der Waals pressure, so that $(\dd\bar{\eta}/\dd \bar{\xi})/\bar{\eta}=\mathrm{const.}$

Finally, we note that this behaviour continues  until the bump reaches the outer limits of the system, i.e.~$\Xmax\approx\Xinf$. The evolution then enters the  very  final  stages  of  the  dynamics in which  gravity  starts to  play  an  important  role (see Appendix \ref{sec:LateTimes}). However, the height of the triple line has  approached its final equilibrium value during late-intermediate times (see fig.~\ref{fig:EvolvingProfiles}e and fig.~\ref{fig:LateIntermediateTimes}a); we therefore turn to discuss when the late-intermediate time behavior ends.

\subsection{\label{sec:FinalTime}Approach to final equilibrium: The role of van der Waals forces and system size}

We have now seen that the film beneath the droplet passes through several different phases as it approaches equilibrium. Of these phases, it is clear that it is the late--intermediate times that dominate the slow approach to equilibrium: figs~\ref{fig:DimpleBumpEvolution} and \ref{fig:LateIntermediateTimes} show that this phase occupies dimensionless times $10^5\lesssim T\lesssim10^8$, corresponding to very long dimensional times. However, fig.~\ref{fig:EvolvingProfiles} also shows that this final approach is heavily influenced by both the size of the system $X_\infty$ and the strength of  van der Waals forces (through the parameter $\AvdW$), even when $\AvdW\ll1$. The dependence on even very weak van der Waals forces is all the more surprising since we saw previously that the final equilibrium state is essentially insensitive to $\AvdW$ unless $\AvdW$ becomes large.  To understand the joint effect of $\AvdW$ and $\Xinf$ in the slow dynamic approach to equilibrium, we continue to neglect the role of gravity, i.e.~taking $\Bo=0$; we shall discuss the effect of $\Bo>0$ shortly.

As might be expected,  the film cannot sense any effect of either van der Waals forces or the system's finite size at early times: figure~\ref{fig:HeightnRot_Late} shows numerical results with various $\AvdW$ and $\Xinf$ collapsing initially. As already discussed,  early times hold for  $t\ll\As^{-4}\tau$, independent of  $\AvdW$. Even in the early-intermediate stages (for which the finite film thickness does play a role),  fig.~\ref{fig:HeightnRot_Late} emphasizes that the initial slowing down in the evolution of the skirt height is independent of $\AvdW$, rather being controlled by the finite thickness of the film and the spreading of the `bump'. The end of this early--intermediate phase does, however, depend on van der Waals forces, taking place when $t=O((\As \AvdW^2)^{-1/3})\tau\propto\mu x_c^{7/3}/(\gamma^{1/3}A^{2/3})$.

Beyond the early and early--intermediate times,  fig.~\ref{fig:LateIntermediateTimes} suggests that the duration of the late-intermediate times is determined by the time taken for the capillary-van der Waals similarity solution to feel the effect of the system size, i.e.~for $\Xinf\propto \AvdW^{3/10}T^{2/5}$ or $T\propto \Xinf^{5/2}\AvdW^{-3/4}\gg1$. While this is correct for systems in which the total volume of liquid limits that in the equilibrium meniscus (lubricant-starved systems in the  equilibrium parlance), it is not correct for lubricant-sated systems where the reservoir of oil available to the skirt is sufficient. In these scenarios, the lateral scale of this similarity solution merely needs to have travelled far enough to suck enough liquid to fill the equilibrium meniscus. With this logic, the effective system size, $\Xinf^{\mathrm{eff}}$, is determined by requiring the volume of liquid in a reservoir of this scale ($\propto1\times\Xinf^{\mathrm{eff}}$ from our non-dimensionalization of film thickness) equal to the ultimate meniscus volume $\propto H_c\Lmen$ where, in the absence of gravity, the lateral scale of the meniscus $\lmen\propto(\gamma h_0^4/A)^{1/2}=x_c\AvdW^{-1/2}$ and so $$\Lmen=\lmen/x_c\propto\AvdW^{-1/2}.$$ 

Since in this limit $h_c\propto x_c$ ($H_c\propto\alpha$), we therefore have that the effective lateral system size is
\begin{equation}
    \Xinf^\mathrm{eff}=\mathrm{min}\left\{\Xinf,\As\AvdW^{-1/2}\right\}.
    \label{eqn:XinfEff}
\end{equation} (Note that, as in the equilibrium problem, we expect the second term in the brackets here to be replaced  by $(\As/\AvdW)^{1/2}$ in an axisymmetric geometry.) The time required to reach equilibrium is thus estimated to be
\begin{equation}
    T_\mathrm{eqm}\propto \begin{cases}
    \AvdW^{-3/4}\Xinf^{5/2},\quad \Xinf\lesssim \As\AvdW^{-1/2}\\
    \As^{5/2}\AvdW^{-2},\quad\Xinf\gtrsim\As\AvdW^{-1/2}
    \end{cases}
    \label{eqn:Teqm}
\end{equation} or in dimensional terms
\begin{equation}
    t_\mathrm{eqm}\propto\begin{cases}
    \frac{\mu \xinf^{5/2}}{A^{3/4}\gamma^{1/4}},\quad\quad \Xinf\lesssim\As\AvdW^{-1/2}\\
    \frac{\mu\gamma x_c^{5/2}h_0^{5/2}}{A^2},\quad \quad \Xinf\gtrsim\As\AvdW^{-1/2}.
    \end{cases}
    \label{eqn:TeqmDim}
\end{equation}
Interestingly, this equilibrium time only depends on the film thickness and the drop size in the lubricant-sated limit and depends on the system size only in the lubricant-starved limit. In both cases, however, the equilibrium time is sensitively dependent on the Hamaker constant $A$. 

\begin{figure}[h!]
    \centering
    \includegraphics[width=9cm]{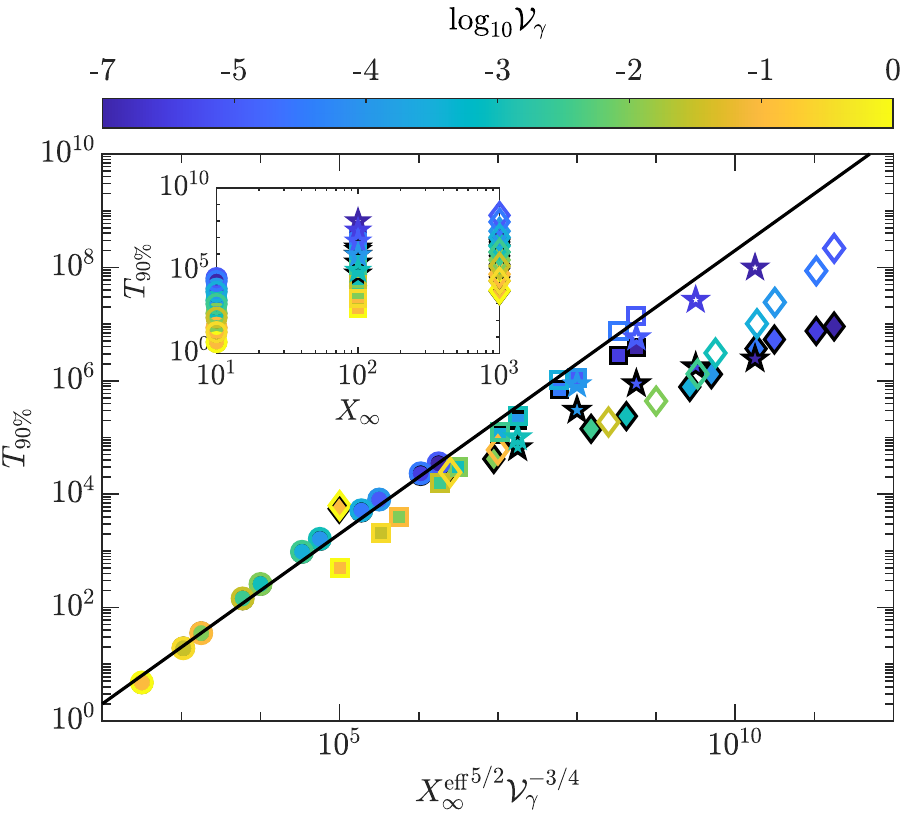}
    \caption{Time for the skirt height to reach 90\% of its final equilibrium value, i.e.~$T_{90\%}$ is defined such that $H(1,T_{90\%})=0.9H_c$. Here, numerical results are shown by points while the scaling of \eqref{eqn:Teqm} with a prefactor 0.02 is shown by the solid line. Marker color is used to encode $\AvdW$ according to the colorbar. Open and filled markers use $\Bo=0$ and $\Bo=10^{-2}$, respectively. Circles, squares, and diamonds denote $\Xinf=10,10^2$ and $10^3$, respectively. All calculations use $\As=10^2$ except pentagrams (for which $\As=25,\Xinf=100$).}    \vspace{0.5cm}
    \label{fig:T90}
\end{figure}

As a test of the scaling laws developed above, we use our numerical simulations to calculate the time taken for the triple line height to reach $90\%$ of its equilibrium value, which we denote by $T_{90\%}$. The dependence of $T_{90\%}$ on the system size is shown in fig.~\ref{fig:T90}. We find good agreement between our numerical results and \eqref{eqn:Teqm} (albeit with a prefactor $\approx1/50$) over a wide range of values of $\Xinf$, $\AvdW$, and $\As$. This analysis was predicated on an equilibrium for the droplet that neglected  gravity, ($\Bo=0$), which is somewhat artificial; we therefore turn now to consider the effects of finite Bond number for all stages of the dynamics.

\subsection{Finite Bond number effects\label{sec:Gravity}}

As a first indication of the important differences introduced by finite $\Bo$ we turn to fig.~\ref{fig:DimpleBumpEvolution}, which shows results for $\As=100,\AvdW=\As^4/10^{12}$ for both $\Bo=0$ and $\Bo=10^{-2}$. In particular, fig.~\ref{fig:DimpleBumpEvolution} shows that differences in the positions and properties of the dimpled and bump regions manifest themselves only at late-intermediate and late times: at early times, the deflection of the interface is small and so the relative contribution of hydrostatic pressure to the flow is minimal. To make this more quantitative, note that in the early and early--intermediate stages of the motion the ratio of typical capillary pressure to typical hydrostatic pressure, $\gamma h_{xx}/(\rho g h)\propto \ell_c^2 /x_{\mathrm{max}}^2$. Since $\Xmax\propto T^{1/4}$ in each case, we therefore find that capillary pressure dominates hydrostatic pressure in these regimes provided that
\begin{equation}
    T\ll \Tgrav= \Bo^{-2};
    \label{eqn:Tgrav}
\end{equation} for the parameters in fig.~\ref{fig:DimpleBumpEvolution}, $\Tgrav=10^4$, agreeing well with the time scale upon which the dashed and solid curves in fig.~\ref{fig:DimpleBumpEvolution} begin to diverge.

For $T\gtrsim\Tgrav$, the bump is flattened out and the dimple propagates away from the triple line region as in the absence of gravity, but both processes take place more quickly with gravity. Similarly, plotting the numerically determined values of $T_{90\%}$ for $\Bo=10^{-2}$ (see filled markers in fig.~\ref{fig:T90}) shows that gravity also accelerates the final approach to equilibrium. To estimate the size of this effect, we first note that the effective system size in this limit must be cut-off by the capillary length $\lc=x_c\Bo^{-1/2}$; hence \eqref{eqn:XinfEff} becomes 
\begin{equation}
\Xinf^\mathrm{eff}=\mathrm{min}\left\{\Xinf,\As(\AvdW+\Bo)^{-1/2}\right\}.
    \label{eqn:XinfEffBoNeq0}
\end{equation} (Again, we expect the second term in the bracket above to be replaced by $\As^{1/2}(\AvdW+\Bo)^{-1/2}$ in axisymmetric situations.) Using this in place of the corresponding result with $\Bo=0$ gives reasonable collapse of the data in fig.~\ref{fig:T90}, particularly for moderate values of the combined parameter ${\Xinf^\mathrm{eff}}^{5/2}\AvdW^{-3/4}$. However, the difference becomes significant as the time scale over which gravity plays a role increases (i.e.~as $T_{90\%}/T_\mathrm{grav}$ increases). Nevertheless, we note that with $\AvdW=10^{-4},\Bo=10^{-2}$, we have $T_\mathrm{90\%}/T_\mathrm{grav}\approx{\Xinf^\mathrm{eff}}^{5/2}/10$ and so might expect \eqref{eqn:Teqm} to give a reasonable estimate of the required time to equilibrium in typical experiments (as long as the system size is not too much greater than the drop). A quick estimation using $\Xinf^\mathrm{eff}=10$ and $\tau\approx1\mathrm{~min}$ gives $t_\mathrm{eqm}\approx4\mathrm{~days}$!

\section{Conclusions\label{sec:Conclusions}}

In this paper, we have studied a model, two-dimensional problem to understand the static and dynamic behaviour of droplets placed on lubricated surfaces. We have focussed on understanding the equilibrium state of the droplet, particularly the size and properties of the skirt that forms around it, and how this skirt forms. To make this problem tractable, we mimicked the effect of the droplet on the lubricating  film as arising from (i) the squeezing provided by its Laplace pressure and (ii) the pulling effect of the droplet--vapor interfacial tension. This `push-and-pull' effect leads to the formation of a skirt qualitatively similar to that often observed experimentally; further, we showed that this model   reproduces previous results for the capillary pressure within the skirt. Crucially, however, this model reveals how the final skirt size depends on the amount of lubricating liquid available in the system. We provided asymptotic relationships for the skirt properties (including height and volume) in the limit of lubricant-starved systems (for which the available lubricant volume is limited) andlubricant-sated systems (for which the available volume is sufficient); interestingly, we found that many experiments are likely to lie  between these two limits. 

\begin{figure}[!htp]
    \centering
    \includegraphics[width=10cm]{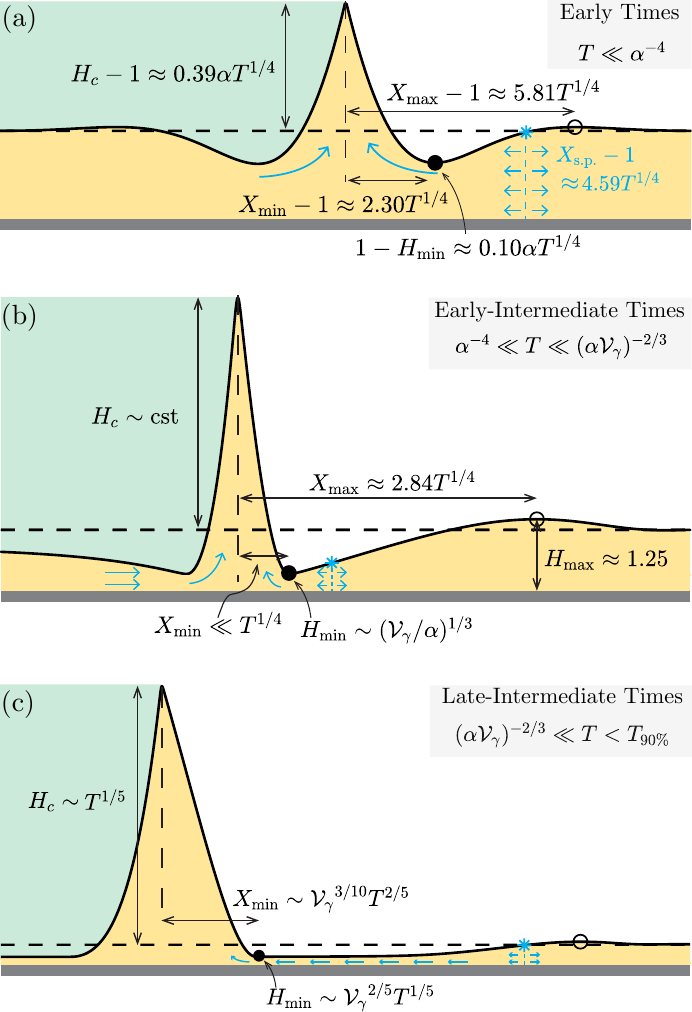}
    \caption{Schematic summary of the asymptotic behaviour of the thin film evolution (the yellow region) as a skirt forms in response to a droplet's capillary pressure (acting in the green region) and the triple line. (a) At early times, capillary pressure is unimportant and a symmetric wedge forms under the action of the triple line, growing according to a similarity solution. (b) At early--intermediate times, van der Waals forces stabilize thin neck regions either side of the skirt; the drainage of the dimple beneath the droplet driven by Laplace pressure is dominant here. (c) At late--intermediate times the skirt is fed mostly by liquid from the remainder of the lubricating film.   The neck and bump regions are highlighted by filled and open circles, respectively;  asterisks indicate the position of the stagnation point between the outer neck and bump regions. The flux of lubricating liquid into the skirt region is illustrated by blue arrows, with the lengths of these arrows representing the relative importance of flow from different regions at different instants of time. }     \vspace{0.5cm}
    \label{fig:RegimesDiagram}
\end{figure}

Our model accounts for attractive van der Waals interactions between the lubricating oil and the substrate --- these act to stabilize the  liquid film beneath the droplet, which would otherwise rupture. For physically relevant values of the strength of  van der Waals forces, their effect on the final equilibrium is slight and limited to determining the thickness of the thin lubricating layer that is trapped beneath the droplet, as suggested previously \cite{Daniel2017}. However, the presence of van der Waals forces leads to a rich evolution of the dynamics of skirt formation through several regimes (see fig.~\ref{fig:RegimesDiagram} for a schematic summary). Crucially, this dynamic role of even moderate van der Waals forces has a significant effect on the time scale taken to reach equilibrium. In particular, we showed that the time scale for the skirt to reach $90\%$ of its equilibrium height is sensitively dependent on the value of the Hamaker constant, see \eqref{eqn:TeqmDim} for example.

A key result of our model is that the evolution of the droplet's skirt towards the ultimate equilibrium is extremely slow: the  early phases of the motion are completed on a time scale $\tau=O(10\mathrm{~s})$ for typical parameters. Crucially, however, the skirt then evolves through several phases, as described in fig.~\ref{fig:RegimesDiagram}, so that on experimentally accessible time scales ($O(10^3\mathrm{~s})$, say), the skirt may only reach $10\%$ of its equilibrium height. As a result, we suggest that most previous experiments with  oil films lubricating smooth substrates are likely to have been in an (admittedly slowly) evolving transient state. While the slow evolution of a droplet on a SLIPS substrate towards equilibrium has been remarked upon before (see \citet{Kreder2018} for example), we believe that the analysis presented here gives new insight into just how slow this motion is and the various features of the problem that cause it. A particular bottleneck seems likely to be the evolution from the early--intermediate to late--intermediate time motions, which we have shown is limited by how quickly the initial dimple beneath the droplet is drained. More evidence  that the drainage of this dimple is a limiting feature of the approach to a true equilibrium is the interference measurements of \citet{Daniel2017}, which show a dimple that is stable over tens of minutes and whose evolution appears to coincide with the evaporation of the droplet itself (see fig.~1c of ref.~\cite{Daniel2017}).

Our model involves several simplifications that should be revisited in future work. Central among these are our assumptions of a two-dimensional droplet sitting above a smooth substrate. In reality, the droplet is likely to be  axisymmetric. At early times this will not impact the analysis presented here (since the lateral size of the skirt region remains small compared to the size of the droplet). However, later the relatively large region of substrate available to drain fluid from to fill the skirt may  increase the speed with which the liquid skirt can grow as well as the system size at which the transition from lubricant-starved to lubricant-sated occurs. (We have made notes of the quantitative changes expected at appropriate points in the main text.) Similarly, when the lubricating oil impregnates a porous coating, the finite thickness of the porous coating is likely to mean that liquid can flow through the porous medium to avoid the `pinch point' beneath the dimple that causes the slow dynamics  discussed here. (Even in the static problem, the axisymmetric and two-dimensional cases can have noticeable differences, as discussed in Appendix \ref{sec:Experiments}.)

Other simplifications made in the development of our push-and-pull model might be expected to have quantitative, rather than qualitative, effects. For example, the discussion of the dynamics presented here has been on the basis of small-slope approximations, which are not self-consistent even in the simplest case that uses the same surface tension for all interfaces. Also inconsistent is the direction of the pulling line force that needs to rotate  to rigorously satisfy horizontal force balance at the triple line. We showed in the static problem that the rotation is often small and the difference between using small and large slopes is only quantitative, even when the system size is unphysically large. However, more general studies that use different surface tensions may need to take the large slope and the rotation of the apparent contact angle (and hence the moving of the triple line) into account. For example, in the case of the droplet becoming  encapsulated in oil (or `cloaked'), the pulling line force would be the sum of the oil-vapor and oil-drop surface tension and such nonlinear effects would be particularly important.  Similarly, we have not accounted for the role of thermal fluctuations or surface roughness in the very thin films that form beneath the droplet during this process. In our view, such intricacies should be included in response to detailed experiments that allow for theoretical predictions to be quantitatively tested. We hope that understanding the possibly long time scale of the droplet evolution in these systems may inspire such experiments.

\begin{acknowledgments}
This project has received funding from the European Union's Horizon 2020 research and innovation programme under the Marie Sklodowska-Curie grant agreement No 886028 (Z.D.) and the Leverhulme Trust through a Philip Leverhulme Prize (D.V.). We also benefited from discussions with Ian Hewitt, Rodrigo Ledesma-Aguilar, Glen McHale and \'{E}lfego Ruiz-Guti\'{e}rrez. 

\end{acknowledgments}

\appendix

\section{Importance of gravity in experiments\label{sec:Experiments}}

\begin{figure}
    \centering
    \includegraphics[width=9cm]{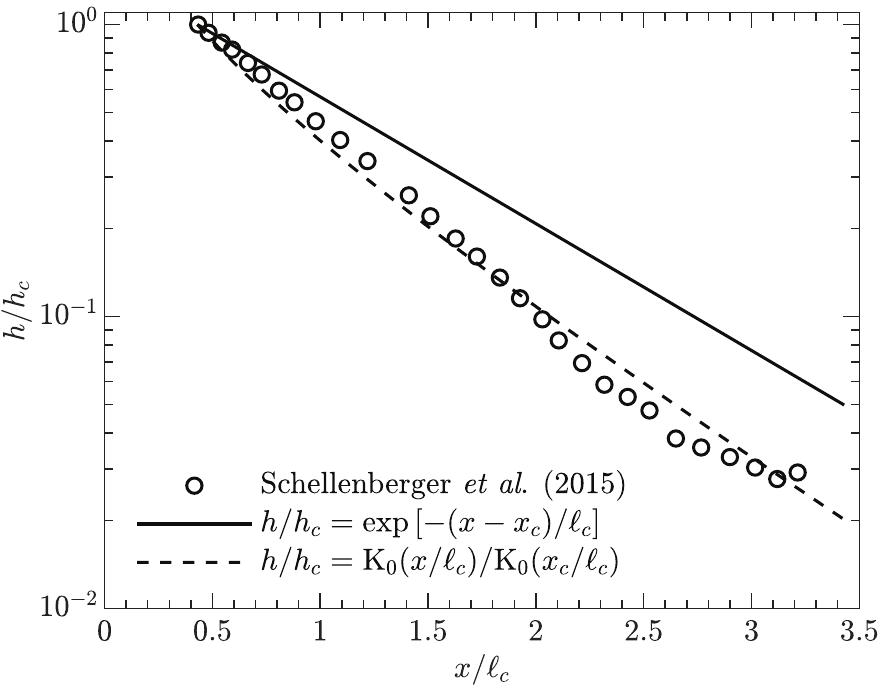}
    \caption{Experimental data for the outer meniscus profile of a droplet on a LIS, captured from  Fig.~5b of \citet{Schellenberger2015}. In these experiments the value of the  capillary length based on the published liquid properties, $\ell_{c}\simeq 0.97 \mathrm{~mm}$, is used. The position of the triple line is taken to be the highest point of the interface visible, which occurs at $r_c=0.43\ell_c$, with the meniscus height there, $h_c\simeq 0.23\mathrm{~mm}$, used to rescale the meniscus height throughout. The prediction from the linearized Laplace--Young equation for a two-dimensional scenario is a pure exponential decay (solid line), while the more detailed, axisymmetric behaviour is given by \eqref{eqn:AxisSymMen}, and is shown by the dashed curve.}    \vspace{0.5cm}
    \label{fig:StaticExp}
\end{figure}

The model we have presented elsewhere in this paper is unusual in that it incorporates both van der Waals forces and hydrostatic pressure. As a single measure of the relative importance of gravity and van der Waals forces, one may introduce the parameter
    \begin{equation}
        \Grav=\Bo/\AvdW=\frac{\rho g h_0^4}{A}.
        \label{eqn:GravDefn}
    \end{equation} The parameter $\Grav$ encodes whether gravity or van der Waals forces are more important in determining how an equilibrium meniscus returns to flat. Interestingly, this number tends to be $O(1)$, and is especially sensitive to the initial thickness of the lubricating film, $h_0$. Typically, therefore, gravity and van der Waals forces place a similar role in the outer reaches of the meniscus. However, closer to the droplet, where $h\gg h_\infty$, we expect that hydrostatic pressure is dominant, and hence that the meniscus shape will approximately follow the solution of the linearized Laplace--Young equation, $\gamma \nabla^2h=\rho g (h-h_\infty)$. In the two-dimensional problems considered here the relevant solution of this equation is an exponential decay over the length scale $\lc=(\gamma/\rho g)^{1/2}$, as seen in the limit of large systems (and late times) already.

Previous experiments \cite{Schellenberger2015} have shown an apparent exponential decay, but over a length scale that is slightly different to the expected capillary $\lc$. This could, perhaps, suggest a stronger force pulling the meniscus down than gravity alone. However, since these experiments are axisymmetric, rather than two-dimensional, the relevant solution of the Laplace--Young equation is
\begin{equation}
h(r)-h_\infty=(h_c-h_\infty)\frac{K_0(r/\ell_c)}{K_0(r_c/\ell_c)},    
\label{eqn:AxisSymMen}
\end{equation} where the triple line is situated at $r=r_c$ and is at height $h_c$ above the plate. (Here $K_0(\cdot)$ is the modified Bessel function of the second kind and zeroth order.)

The experimental data of \citet{Schellenberger2015} allows us to compare the prediction of \eqref{eqn:AxisSymMen} to the two-dimensional solution. Figure \ref{fig:StaticExp} shows the result of digitally capturing the data from fig.~5b of ref.~\cite{Schellenberger2015} and rescaling it according to the expected value of the capillary length $\lc\simeq0.97\mathrm{~mm}$. Plotted in this way, we can see that the axisymmetric meniscus profile of \eqref{eqn:AxisSymMen} gives an excellent account of experimental data \emph{without} any fitting parameters; further, this decay is faster than the pure exponential decay over $\lc$ expected for a two-dimensional meniscus. This agreement validates the approach adopted here of retaining the effect of gravity within our model; this agreement also serves to highlight that the effect of axisymmetry may quantitatively alter the predictions of the two-dimensional analysis presented for  both the static and dynamic problems considered elsewhere in this paper.

\section{Effect of $\AvdW$ on the equilibrium\label{sec:VgEffectStatics}}

\begin{figure}[!h]
    \centering
    \includegraphics[width=8cm]{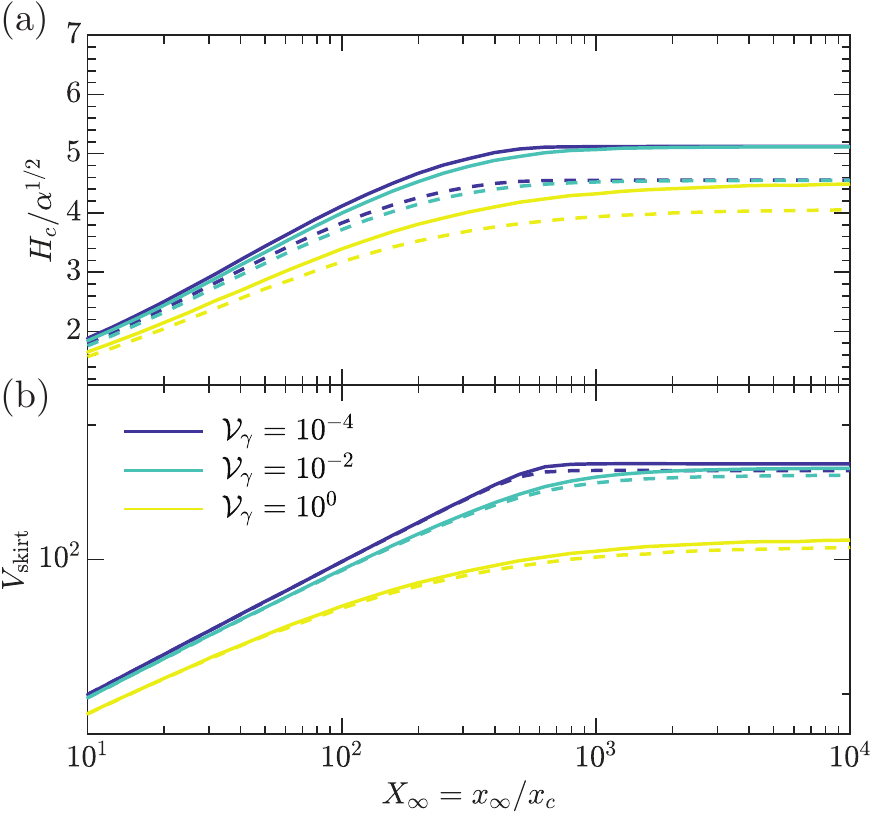}
    \caption{Dependence of (a) the equilibrium skirt height at the triple line and the skirt volume on the strength of van der Waals forces, $\AvdW$, (with $\Bo=10^{-2}$ and $\As=10^{2}$ fixed). Curves with the same colour use the same parameters but are calculated using the fully nonlinear (solid curves) or linearized (dashed curves) curvature.}    \vspace{0.5cm}
    \label{fig:vdWEffectStatics}
\end{figure}

We briefly consider the effect of van der Waals forces on the equilibrium that is established. These forces are important in maintaining a very thin liquid layer on the substrate in the lubricant-starved limit --- see equation \eqref{eq:SmallSkirtAsy}. However, the strength of van der Waals forces does not otherwise enter the asymptotic results \eqref{eq:largeskr} and \eqref{eq:SmallSkirtAsy}. The numerical results shown in Fig.~\ref{fig:StaticProfile}c and fig.~\ref{fig:vdWEffectStatics} show that the effect of the van der Waals parameter on the macroscopic equilibrium properties of the equilibrium setup are very small unless $\AvdW\sim1$, which corresponds to unphysically large values of the Hamaker constant $A$, or ultra-thin liquid films.

\section{Numerical scheme\label{sec:Numerics}}

We only discuss the numerical method for the problem \eqref{eq:EvlEq_Late}-\eqref{eq:BC_Late} here since the results can be directly converted to the problem \eqref{eq:EvlEq_Early}-\eqref{eq:BC_Early} using $\eH=H$, $\eX=\As X$, and $\eT=\As^4T$. We discretize the spatial domain into $N_1$ cells for $[0,1]$ and $N_2$ cells for $[1,\Xinf]$. The  cell widths are non-uniformly distributed in a way that the meshes are highly refined near the triple line. The value of $H$ in the $i$th cell of width $\Delta X_i$, denoted by $H_i$, is evaluated at the mid-point of the cell, while the fluxes, denoted by $Q_{i\pm1/2}$, are evaluated at the end-points. We discretize (\ref{eq:EvlEq_Late}) using central finite differences and obtain a set of ordinary differential equations (ODEs) using the method of lines:
\begin{equation}
    \frac{\dd H_i}{\dd T} = -\frac{1}{\Delta X_i}\left(Q_{i+1/2}-Q_{i-1/2}\right),
    \label{eq:ODE}
\end{equation}
where 
\begin{subequations}
\begin{align}
    Q_{i+1/2}&= -\left(H_{i+1/2}\right)^3\left(\Pi_X-H_{XXX}+\Bo\, H_X\right)_{i+1/2}-3\AvdW H_X|_{i+1/2}/H_{i+1/2},\label{eq:Numerics_Q}\\
    H_{XXX}|_{i+1/2}&=2\left(H_{XX}|_{i+1}-H_{XX}|_{i}\right)/(\Delta X_i+\Delta X_{i+1}),\\
    H_{XX}|_{i}&=\left(H_X|_{i+1/2}-H_X|_{i-1/2}\right)/\Delta X_i,\\
    H_X|_{i+1/2}&=2(H_{i+1}-H_i)/(\Delta X_{i+1}+\Delta X_i),\\
    H_{i+1/2}&=\left(\Delta X_{i+1}H_i+\Delta X_i H_{i+1}\right)/(\Delta X_{i+1}+\Delta X_i),\\
    \Pi_X|_{i+1/2}&=0,\label{eq:Numerics_Pi}
\end{align}
\end{subequations}
for both inner ($1\le i \le N_1$) and outer domain ($1\le i \le N_2$). 

For early-time calculations, the inner and outer menisci are calculated separately so two ghost cells are added on either side of the two domains (8 in total). Besides the four boundary conditions provided by  \eqref{eq:BC_Late}, four matching conditions are used to connect the two menisci at the triple line, which have been discussed in \eqref{eq:matching1}-\eqref{eq:matching4}. 

For late-time calculations, we find using a smoothed step and delta function to be more efficient; in particular, we let
\begin{equation}
    \Pi(X)=\frac{\As}{2}\left(1+\tanh{\frac{1-X}{\varepsilon}}\right)-\frac{\As}{2\varepsilon}\left(\sech{\frac{X-1}{\varepsilon}}\right)^2,
\end{equation}
for $0\le X \le \Xinf$. In this case the evolution of the meshpoints on the two menisci still follows the scheme given by \eqref{eq:ODE} and \eqref{eq:Numerics_Q}-\eqref{eq:Numerics_Pi} in which inner meshes are presented by $i=1,...,N_1$ but the outer meshes are now presented by $i=N_1+1,...,N_1+N_2$. This method could connect the two menisci automatically at the triple line after rewriting (\ref{eq:Numerics_Pi}) as
\begin{equation}
    \Pi_X|_{i\pm1/2}=-\frac{\As}{2\varepsilon^2}\left(\sech{\frac{X_{i\pm1/2}-1}{\varepsilon}}\right)^2\left(\varepsilon+2\tanh{\frac{1-X_{i\pm1/2}}{\varepsilon }}\right),
\end{equation}
for $1\le i \le N_1+N_2$, where $X_{i\pm1/2}$ denotes the position of the right/left end-point of the $i$th cell.

The ODEs (\ref{eq:ODE}) are solved using \textsc{MATLAB}'s built-in solver \texttt{ode15s} and exploiting the system's sparsity as well as complex step differentiation to calculate the Jacobian \cite{shampine2007}. The two methods presented above are identical as $\varepsilon\to0$, but  a finite $\varepsilon$ is used numerically, introducing some errors in early-time calculations (fig. \ref{fig:HeightnRot_Late}). However, it is found that using $\varepsilon=0.01$  provides reasonable accuracy for $T\gtrsim 1$.  The volume of liquid is monitored as an indicator of
the numerical errors accumulated, which we find to remain within 1.5\% of its initial value
for all simulations.  Finally, the presence of repulsive van der Waals forces in the problem ensures a wetting layer remains throughout, so that standard methods are adequate without the need for positivity-preserving schemes \cite{Diez2000,Zhornitskaya2000}. 

\section{Late times}\label{sec:LateTimes}
In this Appendix, we describe the behavior at late times (after the bump has reached the effective edge of the system). We consider separately the cases of zero and finite Bond number.

\subsection{Zero Bond number}

\begin{figure}[htp]
    \centering
    \includegraphics[width=14.5cm]{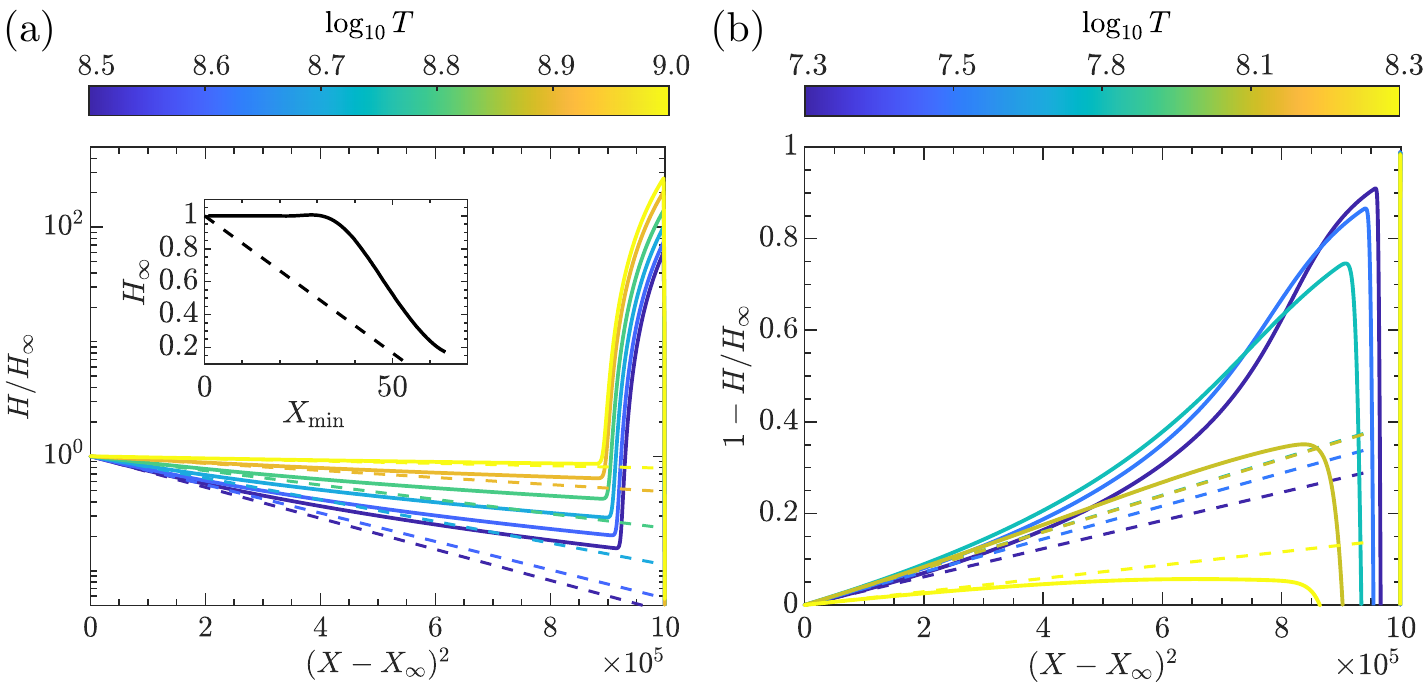}
    \caption{Evolution of the film profile at late times with (a) $\Bo=0$ and (b) $\Bo=10^{-2}$. The colored, dashed lines in (a) and (b) are based on the asymptotic results \eqref{eqn:LateTimesProfile} and \eqref{eqn:LateTimesBoNeq0}, respectively. The inset in (a) shows the relationship between $\Hinf(T)$ and $\Xmin(T)$, with the dashed curve showing the prediction \eqref{eq:LateTimesHinf}. (Here $\Xinf=10^3$, $\alpha=10^2$ and $\AvdW=10^{-4}$ in all calculations.)}    \vspace{0.5cm}
    \label{fig:LateTimes}
\end{figure}

After the bump reaches the outer boundary at $X=X_\infty$, the finite size of the system plays an important role in the further evolution of the system. In particular, the film thickness at the edge, $H_\infty$, starts to decrease, with the lost volume ultimately reaching the skirt, increasing its size  (fig.~\ref{fig:LateTimes}). Since this far-field region is expected to be flat, we neglect the capillary pressure in Reynolds' equation to give
\begin{equation}
   \dot{H}_\infty\approx \frac{\partial}{\partial X}\left(\frac{3\AvdW}{H}\frac{\partial H}{\partial X}\right),
    \label{eq:LateTimes} 
\end{equation} where $\dot{(~)}$ denotes differentiation with respect to time. Equation \eqref{eq:LateTimes} can immediately be integrated twice to give
\begin{equation}
    \frac{\dot{H}_\infty}{2}(X-X_\infty)^2= 3\AvdW\log (H/H_\infty),
    \label{eq:LateTimesFirstInt} 
\end{equation} which may be inverted directly to give the film profile
\begin{equation}
    H = H_\infty \exp\left[\frac{1}{6\AvdW}\dot{H}_\infty(X-\Xinf)^2\right].
    \label{eqn:LateTimesProfile}
\end{equation} Equation \eqref{eqn:LateTimesProfile} leads us to expect a Gaussian-like profile in the outer portion of the film as the system edge is approached ($X\to X_\infty$). This is confirmed by the plots in  fig.~\ref{fig:LateTimes}a where the dashed lines indicate the prediction \eqref{eqn:LateTimesProfile} with the value of $\dot{H}_\infty$ determined separately.

To understand how $\Hinf$ evolves, we consider the flux from the system edge into the skirt. First, however, notice that, at this late stage of the evolution, the skirt height is very close to the final equilibrium, i.e.~$H_c\propto \As$, according to the equilibrium result \eqref{eq:largeskr}. Assuming a parabolic skirt (since $\Bo=0$ and there is little flow), the skirt volume may then be estimated as $\approx \As\Xmin/6$ which must match the change in film volume over the whole system, $\approx(1-\Hinf)\Xinf$, since $1\ll\Xmin\ll\Xinf$; this balance leads to
\begin{equation}
    1-\Hinf(T)\sim\frac{\As\Xmin(T)}{6\Xinf}.
    \label{eq:LateTimesHinf}
\end{equation}
This agrees qualitatively with numerics at very late times (see fig.~\ref{fig:LateTimes}a inset). 

\subsection{Finite Bond number}

ith a finite, non-zero, Bond number, gravity again plays an important role in the very final stages of the dynamics: revisiting the dynamics of the final draining into the meniscus, described by \eqref{eq:EvlEq_Late} with $\partial H/\partial T\approx \dot{H}_\infty$ and analyzing the solution as $X\to\Xinf$ we find that
\begin{equation}
    H\sim H_\infty+\frac{H_\infty \dot{H}_\infty}{2}\frac{(X-X_\infty)^2}{3\AvdW +\Bo\,H_\infty^4},
    \label{eqn:LateTimesBoNeq0}
\end{equation} which is shown as the dashed lines in fig.~\ref{fig:LateTimes}b.
Hence, if $\Bo\gg 3\AvdW\Hinf^{-4}>3\AvdW$, gravity is the relevant restoring force (not van der Waals forces) pulling the interface back towards the far-field liquid surface. Moreover,  in  equilibrium, we have no flux and so the far-field meniscus decay follows:
\begin{equation}
    H-\Hinf\sim \As\exp\left[-(\Bo+3\AvdW\Hinf^{-4})^{1/2}X\right].
\end{equation}
Crucially, for $\Bo\gg 3\AvdW/\Hinf^4$ this is equivalent to the meniscus decay over the capillary length $\ell_c=(\gamma/\rho g)^{1/2}$. While the relative importance of gravity and van der Waals forces is generally governed by $\Grav$ defined in \eqref{eqn:GravDefn}, in the final equilibrium the relevant value of $\Grav$ depends on $\Hinf$ and hence on the size of the system, $\Xinf$; it is therefore possible that a lubricant-starved system with $\Grav\gtrsim1$ may yet have $\Hinf\ll1$ and $\Bo\ll\AvdW/\Hinf^4$.

\bibliography{SLIPS_Biblio}

\end{document}